\documentclass[12pt]{article}

\newtheorem{example}{Example}
\newtheorem{theorem}{Theorem}
\newtheorem{lemma}{Lemma}
\newtheorem{definition}{Definition}

\usepackage{alltt}
\usepackage{graphicx}
\usepackage{color}
\usepackage{nicefrac}
\usepackage{subfigure}
\usepackage{amsmath}
\usepackage{amssymb}
\usepackage{float}
\usepackage{framed}
\usepackage{stmaryrd} 
\usepackage{tikz}
\usepackage{tkz-graph}
\usepackage{lmodern}
\usepackage{fancyvrb, relsize}
\usepackage{listings}
\usepackage{subfigure}
\usepackage{float}
\usepackage{multirow}
\usepackage{paralist}
\usepackage{alltt}
\usepackage{array}
\usepackage{xspace}

\newcommand{\qed}{\blacksquare}
\newcommand{\beginproof}{\bigskip\noindent {\bf Proof:}\ }

\newcommand{\hide}[1]{}
\newcommand{\Inf}[1]{\mathsf{Inf}({#1})}

\newcommand{\Nat}{\mathbb{N}}

\newcommand{\size}[1]{{\mid\!{#1}\!\mid}}

\newcommand{\lift}{{\mathsf {lift}}}
\newcommand{\psol}{{\tt psol}\xspace}

\newcommand{\psolb}{{\tt psolB}\xspace}

\newcommand{\psolq}{{\tt psolQ}\xspace}
\newcommand{\zie}{{\tt zlka}\xspace}
\newcommand{\timeout}{{\tt abo}}
\newcommand{\attr}[3][p]{{{\mathsf{Attr}}_{{#1}}[{#2},{#3}]}}
\newcommand{\nattrword}{{{\mathsf{MAttr}}}}
\newcommand{\nattr}[4][p]{{{\nattrword}_{#1}({#3},{#4})}}
\newcommand{\pnattrword}{{{\mathsf{PMAttr}}}}
\newcommand{\pnattr}[4][p]{{{\pnattrword}_{#1}({#3},{#4})}}
\newcommand{\fattr}[1]{{\mathsf{MA}({#1})}}

\newcommand{\winps}[3]{{{\mathsf{Win}}_{{#1}}[{#2},{#3}]}}

\newcommand{\parts}{{\mathsf \rho}}

\newcommand{\cpred}[2][p]{{\mathsf{cpre}_{#1}({#2})}}
\newcommand{\npredword}{{\mathsf{mpre}}}
\newcommand{\pnpredword}{{\mathsf{pmpre}}}
\newcommand{\npred}[4][p]{{{\npredword}_{#1}({#2}, {#3}, {#4})}}
\newcommand{\pnpred}[4][p]{{{\pnpredword}_{#1}({#2}, {#3}, {#4})}}
\newcommand{\something}{monotone}

\newcommand{\solved}{{\mathcal S}}
\newcommand{\kernel}{{\mathcal K}}
\newcommand{\games}{{\mathcal PG}}

\graphicspath{{./figures/}}
 
\begin{document}

\begin{center}
{\large\bf Fatal Attractors in Parity Games:}
\end{center}
\vspace{-0.7cm}\begin{center}
{\large\bf Building Blocks for Partial Solvers}
\end{center}

\begin{center}
{\it Michael Huth, Jim Huan-Pu Kuo, and Nir Piterman}\ \footnote{Email contact: $\{$m.huth, jimhkuo$\}$@imperial.ac.uk, nir.piterman@leicester.ac.uk.}
\end{center}

\begin{center}
{ 18 May 2014}
\end{center}

\bigskip
{\bf Abstract:}\ Formal methods and verification rely heavily on algorithms that
compute which states of a model satisfy a specified
property. The underlying decision problems are often undecidable or
have prohibitive complexity. Consequently, many algorithms represent
\emph{partial} solvers that may not terminate or report inconclusive results
on some inputs but whose terminating, conclusive outputs are correct. 
It is therefore surprising that partial solvers have
not yet been studied in verification based on parity games, a problem
that is known to be polynomial-time equivalent to the model checking
for modal mu-calculus. We here provide what appears to be 
the first such in-depth technical study by developing the required
foundations for such an approach to solving parity games and by
experimentally validating the utility of these foundations.

Attractors in parity games are a technical device for solving
``alternating'' reachability of given node sets. A well known
solver of parity games~--~Zielonka's algorithm~--~uses such attractor
computations recursively. We here propose new forms of attractors that
are monotone in that they are aware of specific static patterns of
colors encountered in reaching a given node set in alternating
fashion. 
Then we demonstrate how these new forms of attractors can be embedded
within greatest fixed-point computations 
to design solvers of parity games that run in polynomial time but are partial in that they may
not decide the winning status of all nodes in the input game.

Experimental results show that our partial solvers completely solve
benchmarks that were constructed to challenge existing full solvers.
Our partial solvers also have encouraging run times in practice.
For one partial solver we prove that its runtime is at most cubic in
the number of nodes in the parity game, that its
output game is independent of the order in which monotone attractors are
computed, and that it solves all B\"uchi games and weak games.

We then define and study a transformation that converts partial
solvers into more precise partial solvers, and we prove
that this transformation is
sound under very reasonable conditions on the input partial solvers.
Noting that one of our partial solvers meets these conditions, we
apply its transformation on $1.6$ million randomly generated games and
so experimentally validate that the transformation can be very
effective in increasing the precision of partial solvers.

\section{Introduction}\label{sec:introduction}
Formal methods are by now an accepted means of verifying the
correctness of computing artifacts such as
software (e.g.\ \cite{Ball11}), protocol specifications (e.g.\ \cite{Blanchet09},
system architectures (e.g.\ \cite{Bucchiarone08}), behavioral specifications (e.g.\ \cite{Uchitel13}), etc.. Model checking \cite{Clarke81,Queille82} is arguably
one of the success stories of formal methods in this context.
In that approach, one expresses the artifact as a model $M$, the
property to be analyzed as a formula $\phi$, and then can check
whether $M$ satisfies $\phi$ through an automated analysis.
It is well understood \cite{Emerson93,Stirling95} that model checking of an expressive fixed-point
logic, the modal mu-calculus \cite{Kozen83}, is equivalent (within
polynomial time) to the solving of parity games. 

A key obstacle to the effectiveness of model checking and game-based
verification is that the underlying decision problems are typically
undecidable (e.g.\ for many types of infinite-state systems \cite{Esparza01}) or have 
prohibitive complexity (e.g.\ subexponential algorithms for solving
parity games \cite{JPZ06}). Abstraction is seen as an important technique for
addressing these challenges. Program analyses are a good example of
this: a live variable analysis (see e.g.\ \cite{Hankin99}) has as
output statements about
the liveness of program variables at program points. These statements
are of form ``variable $x$ may be live at program point $p$'' and ``$x$ is not live at
$p$'', but that particular live-variable analysis can then not produce
statements of form ``$x$ is definitely live at $p$''.
This limitation trades off undecidability of the underlying decision problem ``is
$x$ live at $p$?'' with the precision of the answer. 

Abstraction of models is another prominent technique for making such
trade-offs \cite{Clarke94,Clarke03}:
instead of verifying that model $M$ satisfies $\phi$, we may abstract $M$ to some model $A$ and
verify $\phi$ on $A$ instead of on $M$. A successful verification on $A$
will then mean that $\phi$ is also verified for $M$. Failure to verify
$\phi$ on $A$ may provide diagnostic evidence for failure of $\phi$
on $M$, but it may alternatively mean that we don't know whether $M$
satisfies $\phi$ or not. So this is again trading off the precision of
the analysis with its complexity: even a verification algorithm that is linear
in the size of $M$ cannot be run on $M$ if the latter is prohibitively
large; but we have to accept that verification on a much smaller
abstraction may be inconclusive for the $M$ in question.

In light of this state of the art in formal verification, we think it
is really surprising that no prior research appears to exist on
algorithms that solve parity games by trading off the
precision of the solution with its complexity. Let us first discuss
parity games at an abstract level before we detail how such a trade
off may work in this setting. 
Mathematically, parity games (see e.g.\ \cite{Zie98}) can be seen as
a representation of the model checking problem for the modal
mu-calculus \cite{Emerson93,Stirling95}, and its exact computational complexity 
has been an open problem for over twenty years now. The decision
problem of which player wins a given node in a given parity game is
known to be in UP$\cap$coUP \cite{Jur98}.

Parity games are infinite, $2$-person, $0$-sum, graph-based games.
Nodes are controlled by different
players (player $0$ and player $1$), are colored with natural numbers, and the winning condition of
plays in the game graph depends on the minimal color occurring in cycles.
A fundamental result about parity games states that these games are
determined \cite{Most91,EJ91,Zie98}: each node in
a parity game is won by exactly one of the players $0$ or
$1$. Moreover, each player has a non-randomized, memoryless strategy
that guarantees her to win from every node that she can indeed win in that game \cite{Most91,EJ91,Zie98}.
The condition for winning a node, however, is an alternation of
existential and universal  quantification. In practice, this means that the maximal color of its
coloring function is the only exponential source for the worst-case
complexity of most parity game solvers, e.g.\ for those in
\cite{Zie98,Jur00,VJ00}. 

We suggest that research on solving parity games may be loosely
grouped into the following different methodological approaches: 
\begin{itemize}
\item design of algorithms that solve all parity games by
construction and that so far all have exponential or
sub-exponential worst-case complexity (e.g.\
\cite{Zie98,Jur00,VJ00,JPZ06}), 

\item restriction of parity games to
classes for which polynomial-time algorithms can be devised as
complete solvers (e.g.\ \cite{BDHK06,Kreutzer12}), 
and 

\item practical improvements to solvers so that they
perform well across benchmarks (e.g.\ \cite{Friedman09}).
\end{itemize}

We here propose a new methodology for researching solutions to
parity games. 
We want to design and evaluate a new
form of ``partial'' parity game solver. These are solvers that are
well defined for all parity games, are guaranteed to run in time
polynomial in the size of the input game, but that may not
solve all games completely~--~i.e.\ for some parity games they may not decide
the winning status of some nodes. 
In this approach, a partial solver has an arbitrary parity game as input and returns 
two things:
\begin{enumerate}
\item a sub-game  of the input game, and 
\item a decision on the winner of nodes from the input game that are not in that sub-game.
\end{enumerate}

In particular, the returned sub-game is empty if,
and only if, the partial solver classified the winners for all input nodes.
The input/output type of our partial solvers clearly relates them to so
called preprocessors that may decide the winner of nodes whose
structure makes such a decision an easy static criterion (e.g.\ in the
elimination of self-loops or dead ends \cite{Friedman09}). 
But we here search for algorithmic building blocks from which we can
construct efficient partial solvers that completely solve a range of
benchmarks of parity games. This ambition sets our work apart from
research on preprocessors but is consistent with it as one can, in principle,
run a partial solver as preprocessor.

The motivation for the study reported in this paper 
is therefore that we want to investigate what
algorithmic building blocks one may create and use for designing
partial solvers that run in polynomial time and work well on
many games, whether there are
interesting subclasses of parity games for which partial solvers
completely solve all game, and whether the insights gained in this
study may lead to an
understanding of whether partial solvers can be
components of more efficient complete solvers. 

We now summarize the main contributions made
in this paper. We present a new form of attractor that can be used in
  fixed-point computations to detect winning nodes for a given player in
  parity games.
Then we propose several designs of partial solvers for parity games
  by using this new attractor within greatest fixed-point computations.
Next, we analyze these partial solvers and show, amongst other things,
  that they work in PTIME and that one of them is independent of the order of
  attractor computation.
We then evaluate these partial solvers against known benchmarks
  and report that these experiments have very encouraging results.
Next, we define a function that transforms partial solvers on that
  class of games to more precise partial solvers. Finally, 
we experimentally evaluate this transformation for the most
  efficient partial solver we proposed in this study and report that
  these experiments are very encouraging.

\emph{Outline of paper.} Section~\ref{sec:preliminaries} contains needed formal
background and fixes notation. Section~\ref{sec:fatalattractors}
introduces the building block of our partial solvers, a new form of
attractor. Some partial solvers based on this attractor are presented
in Section~\ref{sec:partialsolvers}, theoretical results about these
partial solvers are proved in Section~\ref{sec:properties}, and
experimental results for these partial solvers run on benchmarks are
reported and discussed in Section~\ref{sec:results}. The
transformation of partial solvers 
is presented in Section~\ref{sec:transformation}, where we also
analyze and experimentally evaluate this transformation.
We summarize and conclude the paper in
Section~\ref{sec:conclusions}. 

\section{Preliminaries}\label{sec:preliminaries}
We write $\Nat$ for the set $\{0,1,\dots\}$ of natural numbers. 
A parity game $G$ is a tuple $(V,V_0,V_1,E,c)$, where $V$ is a
set of nodes partitioned into possibly empty node sets $V_0$
and $V_1$, with an edge relation $E\subseteq V\times V$ (where for all
$v$ in $V$ there is a $w$ in $V$ with $(v,w)$ in $E$), and a coloring
function $c\colon V\to \Nat$.  In figures, $c(v)$ is written within
nodes $v$, nodes in $V_0$ are depicted
as circles and nodes in $V_1$ as squares. 
For $v$ in $V$, we write $v.E$ for node
set $\{ w\in V\mid (v,w) \in E\}$ of successors of $v$. By abuse of language, we call a
subset $U$ of $V$ a \emph{sub-game} of $G$ if the game graph $(U,E\cap
(U\times U))$ is such that all nodes in $U$ have some successor. We
write $\games$ for the class of all finite parity games $G$, which
includes the parity game with empty node set for our convenience. We
only consider games in $\games$.

Throughout, we write $p$ for one of $0$ or $1$ and $1-p$ for the other
player. 
In a parity game, player $p$ owns the nodes in $V_p$. 
A play from some node $v_0$ results in an infinite play $r =
v_0v_1\dots$ in $(V,E)$ where the player who owns $v_i$ chooses the
successor $v_{i+1}$ such that $(v_i,v_{i+1})$ is in $E$.  
Let $\Inf r$ be the set of colors that occur in $r$ infinitely
often:
\[
\Inf r = \{ k\in\Nat \mid \forall j\in \Nat \colon \exists
i\in \Nat \colon i>j \mbox{ and } k=c(v_i) \}
\]

\noindent Player $0$ wins play $r$ iff $\min \Inf P$ is even; otherwise
player $1$ wins play $r$.

A strategy for player $p$ is a total function $\tau\colon V_p\to V$
such that $(v,\tau(v))$ is in $E$ for all $v\in V_p$. A play $r$ is 
consistent with $\tau$ if each node $v_i$ in $r$ owned by player $p$ 
satisfies $v_{i+1} = \tau(v_i)$. 
It is well known that each parity game is determined:
node set $V$ is the disjoint union of two, possibly empty,
sets $W_0$ and $W_1$, the
winning regions of players $0$ and $1$ (respectively).
Moreover, strategies  $\sigma\colon V_0\to V$
and $\pi\colon V_1\to V$ can be computed such that
\begin{itemize}
\item 
  all plays beginning in $W_0$ and consistent with $\sigma$ are won by
  player $0$; 
  and 
\item 
  all plays beginning in $W_1$ and consistent with $\pi$ are won by
  player $1$. 
\end{itemize}

Solving a parity game means computing such data $(W_0,W_1,\sigma,\pi)$.
\begin{example}
\label{example:pg}
In the parity game $G$ depicted in Figure~\ref{fig:pg},
the winning regions are $W_1 = \{v_3, v_5, v_7\}$ and $W_0 = \{v_0, v_1, v_2, v_4, v_6, v_8, v_9, v_{10}, v_{11}\}$.
Let  $\sigma$ move from $v_2$ to $v_4$, from $v_6$ to $v_8$, from $v_9$ to $v_8$, and from $v_{10}$ to $v_9$. Then $\sigma$ is a winning strategy for player $0$ on $W_0$. And every strategy $\pi$ is winning for player $1$ on $W_1$.
\end{example}

\section{Fatal attractors}\label{sec:fatalattractors}
In this section we define a special type of attractor that is used for
our partial solvers in the next section. 
We start by recalling the normal definition of attractor, and that of
a trap, and then generalize the former to our purposes.
\begin{definition}
Let $X$ be a node set in parity game $G$. For player $p$ in $\{0,1\}$,
set 
\begin{eqnarray}
\cpred {X} & = & \{v \in V_p \mid v.E \cap X \neq \emptyset \} \cup
\{ v\in V_{1-p} \mid v.E \subseteq X \} \label{equ:cpred} \\
\attr{G}{X} & = & \mu Z. (X \cup \cpred{Z}) \label{equ:attr}
\end{eqnarray}

\noindent where $\mu Z.F(Z)$ denotes the least fixed point of a monotone
function $F\colon (2^V,\subseteq)\to (2^V,\subseteq)$.
\end{definition}

The control predecessor of a node set $X$ for $p$ in~(\ref{equ:cpred}) is the set of nodes
from which 
player $p$ can force to get to $X$ in exactly one move.
The attractor for player $p$ to a set $X$ in~(\ref{equ:attr}) is
computed via a least
fixed-point as the set of nodes from
which player $p$ can force the game in zero or more moves to get to the
set $X$.
Dually, a \emph{trap} for player $p$ is a region from which player $p$
cannot escape.

\begin{figure}
\begin{center}
\includegraphics[scale=0.4]{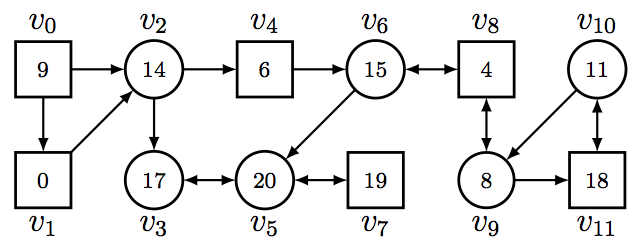}
\end{center}
\caption{
  A parity game: circles denote nodes in $V_0$, squares denote nodes in $V_1$.
  \label{fig:pg}}
\end{figure}

\begin{definition}
Node set $X$ in parity game $G$ is a trap for player $p$ ($p$-trap)
if for all $v\in V_p \cap X$ we have $v.E \subseteq X$ and for all
$v\in V_{1-p}\cap X$ we have $v.E \cap X \neq \emptyset$.
\end{definition}

It is well known that the complement of an attractor for player
$p$ is a $p$-trap and that it is a sub-game. We state this here
formally as a reference:

\begin{theorem}
Given a node set $X$ in a parity game $G$, the set $V\setminus
\attr{G}{X}$ is a $p$-trap and a sub-game of $G$. 
\end{theorem}

We now define a new type of attractor, which 
will be a crucial ingredient in the definition of all our partial
solvers developed in this paper. 
\begin{definition}
Let $A$ and $X$ be node sets in parity game $G$, let
$p$ in $\{ 0,1\}$ be a player, and $c$ a color in $G$.
We set
\begin{eqnarray}
\label{equ:fattr}
\npred{A}{X}{c} &=& 
\{ v\in V_{p}\mid c(v)\geq c \land v.E\cap (A\cup
X)\neq \emptyset\} \cup  \nonumber \\
& & 
\{v\in V_{1-p}\mid c(v) \geq c \land v.E\subseteq A\cup
  X\}\nonumber\\
\nattr{B}{X}{c} &=& \mu Z.\npred{Z}{X}{c}
\end{eqnarray} 
\end{definition}

The \emph{\something} control predecessor $\npred{A}{X}{c}$ 
of node set $A$ for $p$ with target $X$ 
is the set of nodes of color at least $c$ from which player $p$ can
force to get to either $A$ or $X$ in one move.
The \emph{\something} attractor $\nattr{B}{X}{c}$ 
for $p$ with target $X$ 
is the set of nodes from which player $p$ can force the
game in one or more moves to $X$ by only meeting nodes
whose color is at least $c$.
Notice that the target set $X$ is kept external to the attractor.
Thus, if some node $x$ in $X$ is included in
$\nattr{B}{X}{c}$ it is so as it is attracted to $X$
in at least one step.

Our control predecessor and attractor are different from the
``normal'' ones in a few ways.
First, ours take into account the color $c$ 
as a formal parameter. 
They add only nodes that have color at least $c$.
Second, as discussed above, the target set $X$ itself is not included
in the computation by default. 
For example, $\nattr{\emptyset}{X}{c}$ includes states from $X$ only if
they can be attracted to $X$.

We now show the main usage of this new operator by studying how
specific instantiations thereof can compute
so called \emph{fatal attractors}. 
\begin{definition}
Let $X$ be a set of nodes of color $c$, where $p=c\%2$.

\begin{enumerate}
\item For such an $X$ we denote $p$ by $p(X)$ and $c$ by $c(X)$. 
We denote $\nattr{\emptyset}{X}{c}$ by $\fattr{X}$.
If $X=\{x\}$ is a singleton, we denote $\fattr{X}$ by $\fattr{x}$.

\item We say that $\fattr{X}$ is \emph{a fatal attractor} if $X \subseteq
\fattr{X}$.
\end{enumerate}
\end{definition}

We record and prove that fatal attractors $\fattr{X}$ are node sets that are won  
by player $p(X)$ in $G$:
\begin{theorem}
Let $\fattr{X}$ be fatal in parity game $G$. 
Then the attractor strategy for player $p(X)$ on $\fattr{X}$ is
winning for $p(X)$ on $\fattr{X}$ in $G$.
\label{theorem:fatal is winning}
\end{theorem}

\beginproof
The winning strategy is the attractor strategy corresponding to the
least fixed-point computation in $\nattr{\emptyset}{X}{c}$.
First of all, player $p(X)$ can force, from all nodes in $\fattr{X}$,
to reach some node in $X$ in at least one move.
Then, player $p(X)$ can do this 
again from this node in $X$ as $X$ is a subset of $\fattr{X}$. 
At the same time, by definition of $\nattr{\emptyset}{X}{c}$
and $\npred{A}{X}{c}$, the attraction ensures that only colors of value at least $c$ are encountered.
So in plays starting in $\fattr{X}$ and consistent with that strategy,
every visit to a node of parity $1-p(X)$ is followed later by a visit
to a node of color $c(X)$.
It follows that in an infinite play consistent with this strategy and starting in $\fattr{X}$, the
minimal color to be visited infinitely often is $c$~--~which is of $p$'s
parity. $\qed$\bigskip

Let us consider the case when $X$ is a singleton $\{k\}$ and
$\fattr{k}$ is not fatal. We show that, under a certain condition, we
can remove an edge from $G$ without changing the set of winning
strategies or winning regions of either player:
\begin{lemma}
Let $\fattr{k}$ be not fatal for node $k$.
Then we may remove edge $(k,w)$ in $E$ if
$w$ is in $\fattr{k}$, without changing winning regions of parity
game $G$.
\label{lemma:edge removal is sound}
\end{lemma}

\beginproof
Suppose that there is an edge $(k,w)$ in $E$ with $w$ in
$\fattr{k}$. 
We show that this edge cannot be part of a winning strategy (of
either player) in $G$.
Since $\fattr{k}$ is not fatal, $k$ must be in $V_{1-p(k)}$
and so is controlled by player $1-p(k)$. 
But if that player were to move from $k$ to $w$ in a memoryless
strategy, player $p(k)$ could then attract the play from $w$ back to
$k$ without visiting colors of parity $1-p(k)$ and smaller
than $c(k)$, since $w$ is in $\fattr{k}$. 
And, by the existence of memoryless winning strategies \cite{EJ91},
this would ensure that the play is won by player
$p(k)$ as the minimal infinitely occurring color would have parity
$p(k)$.
 $\qed$\bigskip

\begin{example}
For $G$ in Figure~\ref{fig:pg}, the only colors $k$ for
which $\fattr{k}$ is fatal are $4$ and $8$: monotone attractor $\fattr{4}$ equals $\{v_2,
v_4, v_6, v_8, v_9, v_{10}, v_{11}\}$ and monotone attractor $\fattr{8}$ equals $\{v_9,
v_{10}, v_{11}\}$. In particular, $\fattr{8}$ is contained in
$\fattr{4}$ and nodes $v_1$ and $v_0$ are attracted to $\fattr{4}$ in
$G$ by player $0$. And $v_{11}$ is in $\fattr{11}$ (but the node of color 11, $v_{10}$, is not), so edge $(v_{10},v_{11})$
may be removed.
\end{example}

\section{Partial solvers}\label{sec:partialsolvers}

We can use the above definitions and results to define partial solvers
next. In doing so, we will also prove their
soundness. \emph{Throughout this paper, pseudo-code of partial solvers
  will not show the
  routine code for accumulating detected winning regions and their
  corresponding winning strategies~--~their
  nature will be clear from our discussions and soundness proofs.}


\subsection{Partial solver \psol}
Figure~\ref{fig:partialsolver} shows the pseudocode of a partial
solver, named \psol, based on $\fattr{X}$ for singleton sets $X$.
Solver \psol explores the parity game $G$ in descending color
ordering. 
For each node $k$, it constructs $\fattr{k}$, and aims to do one of
two things: 

\begin{itemize}
\item 
  If node $k$ is in $\fattr{k}$, then $\fattr{k}$ is fatal for
  player $1-p(k)$, thus node set $\attr[p(k)]G{\fattr{k}}$ is a winning
  region of player $p(k)$, and removed from $G$. 
\item 
  If node $k$ is not in $\fattr{k}$, and there is a $(k, w)$
  in $E$
  where $w$ is in $\fattr{k}$, all such edges $(k, w)$ are removed from $E$ and 
  the iteration continues. 
\end{itemize}

\begin{figure}[bt]
\begin{center}
{\small
\begin{alltt}
psol(\(G=(V,V\sb{0},V\sb{1},\!E,c)\)) \{
  for (\(k\in\!\! V\) in descending color ordering \(c(k)\)) \{
    if (\(k\in{\fattr{k}}\)) \{ return psol(\(G \!\!\setminus\!\! {\attr[p(k)]{G}{\fattr{k}}}\)) \} 
    if (\(\exists\!\! (k,w)\in\!\! E\colon\!\! w\in\! \fattr{k}\))
    \(\{ G = G\setminus \!\!\!\!\!\! \{(k,w)\in \!\!E\mid w\in\!\! \fattr{k}\} \}\) 
  \}
  return \(G\)
\}
\end{alltt}
}
\end{center}
\caption{Partial solver \psol\ based on detection of fatal
  attractors $\fattr{k}$ and fatal moves.\label{fig:partialsolver}} 
\end{figure}

If for no $k$ in $V$ attractor $\fattr{k}$ is fatal, game $G$ is
returned as is~--~empty if $\psol$ solves $G$ completely.

\begin{example}
In a run of \psol on $G$ from Figure~\ref{fig:pg}, there is no effect
for colors larger than $11$. For $c=11$, \psol removes edge $(v_{10},
v_{11})$ as $v_{11}$ is in the monotone attractor $\fattr{11}$. The next effect is for $c=8$, when the fatal attractor $\fattr{8} = \{v_9, v_{10}, v_{11}\}$ is detected and removed from $G$ (the previous edge removal did not cause the attractor to be fatal).
On the remaining game, the next effect occurs when $c=4$, and when the
fatal attractor $\fattr{4}$ is $\{v_2, v_4, v_6, v_8\}$ in that
remaining game. As player $0$ can attract $v_0$ and $v_1$ to this as
well, all these nodes are removed and the remaining game has node set
$\{v_3, v_5, v_7\}$. As there is no more effect of \psol on that
remaining game, it is returned as the output of \psol's run.
\end{example}


\subsection{Partial solver \psolb}
Figure~\ref{fig:buchi} shows the pseudocode of another partial solver, named
\psolb~--~the ``B'' suggesting a relationship to ``B\"uchi''. This
partial solver is based on $\fattr{X}$, where $X$ is a set of nodes of the same
color. 
This time, the operator $\fattr{X}$ is used within a greatest
fixed-point in order to discover the largest set of nodes of a certain
color that can be (fatally) attracted to itself.
Accordingly, the greatest fixed-point starts from all the nodes of a
certain color and gradually removes those that cannot be attracted to
the same color.
When the fixed-point stabilizes, it includes the set of nodes of the
given color that can be (fatally) attracted to itself.
This node set can be removed (as a winning region for player $d\%2$) and the residual game
analyzed recursively. 
As before, the colors are explored in descending order. 

\begin{figure}[bt]
\begin{center}
{\small
\begin{alltt}
psolB(\(G=(V,V\sb{0},V\sb{1},\!E,c)\)) \{
  for (colors \(d\) in descending ordering) \{
    \(X = \{ v\ in V\mid c(v) = d\, \}\);
    cache = \(\{\}\);
    while (\(X\neq \{\}\) \&\& \(X\neq \) cache) \{
      cache = \(X\);
      if (\(X\subseteq {\fattr{X}}\)) \{ return psolB(\(G \setminus {\attr[d\%2]{G}{\fattr{X}}}\))  
      \} else \{ \(X\) = \(X\cap\! {\fattr{X}}\); \}
    \}
  \}
  return \(G\)
\}
\end{alltt}
}
\end{center}
\caption{Partial solver \psolb.\label{fig:buchi}} 
\end{figure}

We make two observations. First, if we were to
replace the recursive calls in \psolb with the removal of the winning
region from $G$ and a continuation of the iteration, we would get an
implementation that discovers less fatal attractors~--~we confirmed
this experimentally.
Second, edge removal in \psol relies on the set $X$ being a singleton. 
A similar removal could be achieved in \psolb when the  size of $X$
is reduced by one (in the operation $X=X\cap \fattr{X}$).
Indeed, in such a case the removed node would not be removed
and the current value of $X$ be realized as fatal.
We have not tested this edge removal approach experimentally for this variant of \psolb.

\begin{example}
A run of \psolb on $G$ from Figure~\ref{fig:pg} has the same effect as
the one for \psol, except that \psolb does not remove edge $(v_{10}, v_{11})$ when $c=11$.
\end{example}

\begin{figure}[htb]
\begin{center}
\subfigure[]{
\includegraphics[scale=0.4]{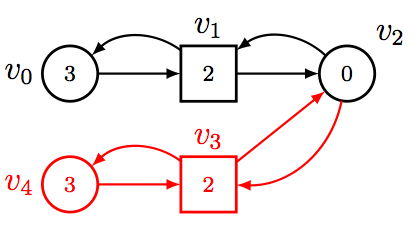}
}
\qquad
\subfigure[]{
\hide{
\begin{tikzpicture}[scale=1, transform shape]
\Vertex[x=0.0pt,y=0.0pt,L=$v_{0}$,LabelOut=true,Ldist=0pt,Lpos=180,style={color=white,text=black}]{s0ext}
\Vertex[x=0.0pt,y=0.0pt,L=$0$,style={font=\scriptsize, minimum size=20pt}]{v0}
\Vertex[x=50.0pt,y=0pt,L=$v_{1}$,LabelOut=true,Ldist=0pt,Lpos=90,style={color=white,text=black}]{s1ext}
\Vertex[x=50.0pt,y=0pt,L=$1$,style={font=\scriptsize, shape=rectangle,minimum height=20pt, minimum width=20pt}]{v1}
\Vertex[x=100.0pt,y=0.0pt,L=$v_{2}$,LabelOut=true,Ldist=0pt,Lpos=0,style={color=white,text=black}]{s2ext}
\Vertex[x=100.0pt,y=0.0pt,L=$0$,style={font=\scriptsize, minimum size=20pt}]{v2}
\Edge[style={<->,>=latex}](v0)(v1)
\Edge[style={<->,>=latex}](v2)(v1)
\end{tikzpicture}
}
\includegraphics[scale=0.4]{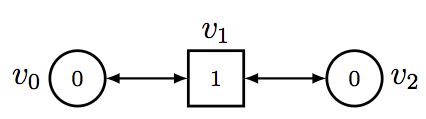}
}
\end{center}
\caption{Subfigure (a): a game that \psolb cannot solve at all and that \psol
  solves completely. Subfigure (b): a game that \psol cannot solve at
  all but that \psolb solves completely.
  \label{fig:incomparable}}
\end{figure}

A way of comparing partial solvers $P_1$ and $P_2$ is to say that
$P_1\leq P_2$ if, and only if, for all parity games $G$ the set of
nodes in the output sub-game $P_1(G)$ is a subset of the set of nodes
of the output sub-game $P_2(G)$. The next example shows that \psol and \psolb are
incomparable for this intentional pre-order over partial
solvers:
\begin{example}
Consider the game $G$ in Figure~\ref{fig:incomparable}(a).
Partial solver \psolb decides no nodes in this game since the
monotone attractors it computes are empty for all colors of $G_1$.
But \psol detects for $k=v_1$ that $v_0$ is in the monotone attractor of
$k$ and that $v_1$ is not. Therefore, it removes edge $(v_1,v_0)$ from
$G_1$. When it comes to evaluating $k=v_2$, it now detects that $v_2$
is in its monotone attractor and so this fatal attractor decides
$\{v_0,v_1,v_3\}$. 
The same process repeats for $v_3$.
We note that when \psolb computes the monotone attractor of
$\{v_1,v_3\}$ both nodes are removed from the attractor
simultaneously.
Thus, our optimization of \psolb that tries to remove edges when the
size of the set decreases by $1$ does not apply here.

Now consider game $G'$ in Figure~\ref{fig:incomparable}(b).
Then all monotone attractors that \psol computes are empty and so
it solves no nodes. But running \psolb on $G'$ now decides all
nodes since it detects for $d = 0$ and $X = \{v_0,v_2\}$ a fatal
attractor for all nodes.
\end{example}
%

Let us introduce some notation for the regions of nodes that are
decided by partial solves.
\begin{definition}
Let $G$ be a parity game, $\parts$ a partial solver, and $p$ in $\{0,1\}$ a player in
$G$. Then $\winps {\parts}Gs$ denotes the set of nodes in $G$ that $\parts$
classifies as being won by player $p$ in $G$.
\end{definition}

Next, we state that both \psolb and \psol are sound partial solver.
\begin{theorem}
\label{theorem:psol soundness}
The partial solvers \psolb and \psol are sound: let $G$ be a parity game.
Then $\winps {\psol}G0$ and $\winps {\psol}G0$ are contained in the winning region of player $0$ in $G$,
and $\winps {\psolb}G1$ and $\winps {\psolb}G1$ are contained in the winning region of player $1$ in $G$. 
\end{theorem}

\beginproof
\begin{enumerate}
\item We prove the claim for \psolb first. Let $G$ be a parity game.
In Theorem~\ref{theorem:fatal is winning}, we have proved that
$\fattr{X}$ is winning for player $p(X)$ if $X$ is a subset of $\fattr{X}$.
For every color $d$ in
$G$, the for-loop in \psolb constructs $\fattr X$ where all nodes in
$X$ have color $d$.
If $X$ is a subset of $\fattr X$, then $\fattr X$ is identified as a winning
region (for player $d\%2$) and its normal $d\%2$ attractor in $G$ is
therefore removed from $G$, and this is the only code location where $G$ is
modified.

\item We prove the claim for \psol next.
In Figure~\ref{fig:partialsolver}, \psol only returns (not explicitly
shown) $\attr[p(k)]G{\fattr{k}}$ as a node set classified to be won by
player $p(k)$ whenever $\fattr{k}$
is fatal. Theorem~\ref{theorem:fatal is winning} shows that
these regions are winning for player $p(k)$.
Lemma~\ref{lemma:edge removal is sound} shows edge removal does not
alter the winning strategies. Since these are the only two code
locations where $G$ is modified, the winning regions
detected in \psol are correct.
\end{enumerate}
 $\qed$\bigskip


\subsection{Partial solver \psolq}

It seems that \psolb is more general than \psol in that if there is a
singleton $X$ with $X \subseteq \fattr{X}$ then \psolb will
discover this as well.
However, the requirement to attract to a single node seems too strong.
Solver \psolb removes this restriction and allows to attract to more
than one node, albeit of the same color.
Now we design a partial solver \psolq 
that can attract to a set of
nodes of more than one color~--~the ``Q'' is our code name for this
``Q''uantified layer of colors of the same parity.
Solver \psolq allows to combine attraction to multiple colors by
adding them gradually and taking care to ``fix'' visits to nodes of
 opposite parity.

We extend the definition of $\npredword$ and
$\nattrword$ to allow inclusion of more (safe) nodes when collecting
nodes in the attractor.
\begin{definition}
Let $A$ and $X$ be node sets in parity game $G$, let $p$ in $\{0,1\}$
be a player, and $c$ a color in $G$.
We set
\begin{eqnarray}
\pnpred{A}{X}{c} &=& 
\{ v\in V_{p}\mid (c(v)\geq c \vee v\in X) \land v.E\cap (A\cup
X)\neq \emptyset\} \cup  \nonumber \\
& & 
\{v\in V_{1-p}\mid (c(v) \geq c \vee v\in X) \land v.E\subseteq A\cup
  X\} \label{equ:pmpre} \\
\pnattr{B}{X}{c} &=& \mu Z.\pnpred{Z}{X}{c} \label{equ:pfattr}
\end{eqnarray} 
\end{definition}

The \emph{permissive \something} predecessor in~(\ref{equ:pmpre}) adds to the {\something}
predecessor also nodes that are in $X$ itself even if their color is
lower than $c$, i.e., they violate the monotonicity requirement.
The \emph{permissive \something} attractor in~(\ref{equ:pfattr}) then uses the permissive
predecessor instead of the simpler predecessor.
This is used for two purposes. 
First, when the set $X$ includes nodes
of multiple colors~--~some of them lower than $c$.
Then, inclusion of nodes from $X$ does not destroy the properties of
fatal attraction.
Second, increasing the set $X$ of target nodes allows to include the
previous target as set of ``permissible'' nodes.
This creates a layered structure of attractors.

\begin{figure}[bt]
\begin{center}
{\small
\begin{alltt}
layeredAttr(\(G\),\(p\),\(X\)) \{ // PRE-CONDITION: all nodes in \(X\) have parity \(p\)
   \(A\) = \(\{\}\);
   \(b\) = max\(\{c(v)\,\mid\,v\,\in\,X\}\);
   for (\(d\) = \(p\) up to \(b\) in increments of 2) \{
      \(Y\) = \(\{v\,\in\,X\,\mid\,c(v)\,\leq\,d\}\);
      \(A\) = \(\pnattr{\emptyset}{A\,\cup\,Y}{d}\);
   \}
   return \(A\);
\}
\end{alltt}
}

\smallskip
{\small
\begin{alltt}
psolQ(\(G=(V,V\sb{0},V\sb{1},\!E,c)\)) \{
   for (colors \(b\) in ascending order) \{
      \(X\) = \{\(v\,\in\,V\,\mid\,c(v)\,\leq\,b\,\wedge\,c(v)\%2=b\%2\)\};
      cache = \(\{\}\);
      while (\(X \neq \{\}\) \&\& \(X\neq \)cache) \{
         cache = \(X\);
         \(W\) = layeredAttr(\(G\),\(b\%2\),\(X\));
         if (\(X\,\subseteq\,W\)) \{ return psolQ(\(G\,\setminus\,{\attr[b\%2]{G}{W}}\)); 
         \} else \{ \(X\) = \(X\,\cap\,W\); \}
      \}
   \}
   return \(G\);
\}
\end{alltt}
}
\end{center}
\caption{Operator {\tt layeredAttr}($G,p,X$) and partial solver \psolq.\label{fig:fixedpointq}}
\end{figure}

We use the permissive attractor to define \psolq.
Figure~\ref{fig:fixedpointq} presents the pseudo-code of operator
{\tt layeredAttr}($G,p,X$). It is an attractor that combines
attraction to nodes of multiple color.
It takes a set $X$ of colors of the same parity $p$.
It considers increasing subsets of $X$ with more and more colors and
tries to attract fatally to them.
It starts from a set $Y_p$ of nodes of parity $p$ with color $p$
and computes $\fattr{Y_p}$. 
At this stage, the difference between $\pnpredword$ and $\npredword$
does not apply as $Y_p$ contains nodes of only one color and $A$ is
empty.
Then, instead of stopping as before, it continues to accumulate more
nodes.
It creates the set $Y_{p+2}$ of the nodes of parity $p$ with color $p$ or
$p+2$.
Then, $\pnattr{\emptyset}{A \cup Y_{p+2}}{p+2}$ includes all the previous nodes
in $A$ (as all nodes in $A$ are now permissible) and all nodes that
can be attracted to them or to $Y_{p+2}$ through nodes of color at least
$p+2$.
This way, even if nodes of a color lower than $p+2$ are included they
will be ensured to be either in the previous attractor or of the right
parity.
Then $Y$ is increased again to include some more nodes of $p$'s
parity. 
This process continues until it includes all nodes in $X$.

This layered attractor may also be fatal:

\begin{definition}
We say that {\tt layeredAttr}($G,p,X$) is fatal if
$X$ is a subset of $\mbox{{\tt layeredAttr}(}G,p,X\mbox{)}$.
\end{definition}

As before, fatal layered attractors  {\tt layeredAttr}($G,p,X$) are won by player $p$ in $G$. 
The winning strategy is more complicated as it has to take into
account the number of iterations in the for loop in which a node was
first discovered.
Every node in {\tt layeredAttr}($G,p,X$) belongs to a 
layer corresponding to a maximal color $d$. 
From a node in layer $d$, player $p$ can force to reach some node in
$Y_d\subseteq X$ or some node in a lower layer $d'$.
As the number of layers is finite, eventually some node in $X$ is
reached.
When reaching $X$, player $p$ can attract to $X$ in the same layered
fashion again as $X$ is a subset of {\tt layeredAttr}($G,p,X$).
Along the way, while attracting through layer $d$ we are 
ensured that only colors at least $d$ or of a lower layer are
encountered. 
So in plays starting in {\tt layeredAttr}($G,p,X$) and consistent
with that strategy, every visit to a node of parity $1-p$ is followed
later by a visit to a node of parity $p$ of lower color.
We formally state the soundness of \psolq and extend the above
argument to a detailed soundness proof:
\begin{theorem}
Let {\tt layeredAttr}($G,p,X$) be fatal in parity game $G$.
Then the layered attractor strategy for player $p$ on {\tt
  layeredAttr}($G,p,X$) is winning for $p$ on {\tt
  layeredAttr}($G,p,X$) in $G$.
\end{theorem}

\beginproof
We show that if $X\subseteq\mbox{{\tt layeredAttr}}(G,p,X)$ is winning
for $p$. Without loss of generality, $p$ equals $0$.

By assumption all nodes in $X$ have parity $p$.
Let $b$ be the maximal color in $b$.
Let $A_{d}$ be an enumeration of the sets A computed by the
instruction $A=\pnattr{}{A\cup Y}{d}$.
It follows that $A_0=\pnattr{}{X_0}{0}$ and
$A_{d}=\pnattr{}{A_{d-2}\cup X_{d}}{d}$, where $X_d$ is the set of
nodes in $X$ of color at most $d$.
It follows that $A_b$ is the value of 
$\mbox{{\tt layeredAttr}}(G,p,X)$. 
Note in the {\tt layeredAttr} $b$ is a constant.
Let $A_{d,i}$ be a partition of $A_d$ according to the iteration
number in computing $\pnattr{}{A_{d-2}\cup X_d}{d}$.
For every node $v$ in $A$, let $r(v)=(d,i)$ be minimal in the
lexicographic order such that $v$ is in $A_{d,i}$.
We choose the strategy that selects the successor with minimal
$r$ according to the same lexicographic order.  

Consider an infinite play starting in $A_b$ in which player 0 follows 
this strategy.
First, we show that the play remains in $A_b$ forever. 
Indeed, if $r(v)=(0,1)$ then all successors of $v$ 
(if $v\in V_{1=b\%2}$) or some successor of $v$ (if $v\in V_{b\%2}$)
are in $X$ and $X\subseteq A_d$.
If $r(v)=(d,i)>(0,1)$ then all successors of $v$ (if $v$ $\in
V_{1-b\%2}$) or some successor of $v$ 
(if $v\in V_{b\%2}$) are/is either in 
$X$, or in $A_{d',i'}$ for some 
$(d',i')<r(v)$.

Second, we show that the play is winning for player 0. 
Consider an $odd$ colored node $v_0$ appearing in the play.
Let $v_0,v_1,\ldots$ be an enumeration of the nodes in the play
starting from $v_0$.
By definition, $v_0$ is in $A_{d_0,i_0}$ for some $(d_0,i_0)$, and clearly,
$c(v_0)> d_0$. 
We have to show that this play visits some even color that is at most
$d_0$. 
By construction, $v_1$ is either in \{$v \in X$ $\mid$ $c(v)$
$\le$ $d_0$\}, 
which implies that its color is even and smaller than $c(v_0)$, or in
$A_{d_1,i_1}$ for 
some $(d_1,i_1)<(d_0,i_0)$.
In this case, the obligation to visit an even color at most $d_0$ is
passed to $v_1$.
We strengthen the obligation to visit an even color at most $d_1$.
Continuing this way, the play must reach $X$ with a lower color than
that of $v_0$ by well-founded induction. 
 $\qed$\bigskip

Pseudo-code of solver \psolq\ is
also shown in Figure~\ref{fig:fixedpointq}:
\psolq prepares increasing sets of nodes $X$ of the same color and calls
{\tt layeredAttr} within a greatest fixed-point.
For a set $X$, the greatest fixed-point attempts to discover the
largest set of nodes within $X$ that can be fatally attracted to
itself (in a layered fashion).
Accordingly, the greatest fixed-point starts from all the nodes in $X$
and gradually removes those that cannot be attracted to $X$.
When the fixed-point stabilizes, it includes a set of nodes of the
same parity that can be attracted to itself.
These are removed (along with the normal attractor to them) and the
residual game is analyzed recursively. 

We note that the first two iterations of \psolq are equivalent to calling
\psolb on colors $0$ and $1$.
Then, every iteration of \psolq extends the number of colors
considered.
In particular, in the last two iterations of \psolq the value of $b$
is the maximal possible value of the appropriate parity.
It follows that the sets $X$ defined in these last
two iterations include all nodes of the given parity.
These last two computations of greatest fixed-points are the most
general and subsume all previous greatest fixed-point computations.
We discuss in Section~\ref{sec:results} why we increase the bound $b$
gradually and do not consider these two iterations alone.

\begin{figure}
\begin{center}
\includegraphics[scale=0.4]{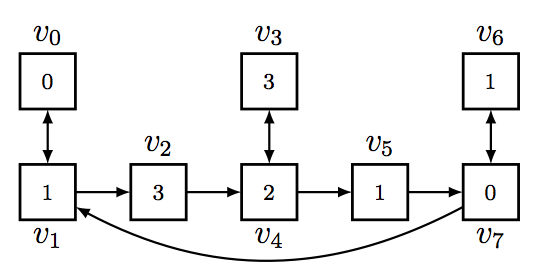}
\end{center}
\caption{
  A 1-player parity game modified by neither \psol, \psolb nor \psolq.
\label{fig:pgtwo}}
\end{figure}

\begin{example}
The run of \psolq on $G$ from Figure~\ref{fig:pg} finds a fatal
attractor for bound $b=4$, which removes all nodes except $v_3, v_5$,
and $v_7$. For $b=19$, it realizes that these nodes are won by player
$1$, and outputs the empty game. That \psolq is a partial solver can
be seen in Figure~\ref{fig:pgtwo}, which depicts a game that is not
modified at all by \psolq and so is returned as is.
\end{example}

\section{Properties of our partial solvers}\label{sec:properties}

We already proved that our partial solvers are sound. Now we want to
investigate additional properties of these partial solvers, looking
first at their computational complexity.

\subsection{Computational Complexity}
We record that the partial solvers we designed above all run in
polynomial time in the size of their input game.
\begin{theorem}
Let $G$ be a parity game with node set $V$, edge set $E$, and $\size
c$ the number of its different colors.
\label{theorem:complexity}
\begin{enumerate}
\item The running time for partial solvers \psol and \psolb is in $O({\size V}^2\cdot
\size E)$.
\item The partial solvers \psol and \psolb can be implemented to run in time $O({\size V}^3)$.
\item The partial solver \psolq runs in time $O({\size V}^2\!\cdot\!
  \size E \!\cdot\! \size c)$.
\end{enumerate}
\end{theorem}

\beginproof
\begin{enumerate}
\item To see that the running time for \psol is in $O({\size V}^2\cdot
\size E)$, note that all nodes have at least
one successor in $G$ and so $\size V\leq \size E$. The
computation of the attractor $\fattr{k}$ in linear in the number of
edges and so in $O(\size E)$. Each call of \psol will compute at most
$\size V$ many such attractors. In the worst case, there are $\size V$
many recursive calls. In summary, the running time is bound by
$O(\size E\cdot \size V\cdot \size V)$ as claimed.

To see that \psolb also has running time in $O({\size V}^2\cdot
\size E)$, recall that we may compute $\fattr{X}$ in time linear in
$\size{E}$. Second, node set $V$ is partitioned into sets of nodes of a specific
color, and so \psolb can do at most $\size{V}$ many computations within
the body of \psolb before and if a recursive call happens.

\item The claim that \psol and \psolb can be implemented to run in
$O({\size V}^3)$ essentially reduces to showing that we can, in linear
time, transform and reduce each computation of $\fattr{X}$ to the
solution of a Buchi game. This is so since Buchi games can be solved in
time $O({\size V}^2)$, as shown in~\cite{ChatterjeeH12}. 
Indeed, let $c$ denote $c(X)$, $p$ denote $p(X)$, and let $G[{\geq}c]$
denote the game obtained from $G$ by doing the following in the
prescribed order.
\medskip
\begin{enumerate}
\item
Remove from $G$ all nodes of color less than $c$, as well as all of
their incoming and outgoing edges.
\item
Add to $G$ a sink node 
that has a self loop.
\item
Every node in $V_p$ not removed in the first step but where
all of its successors were removed gets an edge to
the new sink node.
\item
Every node in $V_{1-p}$ not removed in the first step but that had one of its
successors removed gets an edge to the new sink
node as well.
\item
If $p=1$, then we swap ownership of all remaining nodes: player $0$ nodes
become player $1$ nodes, and vice versa.
\item
Finally, we color every node in $X$ by $0$ and all other nodes (including the new
sink state) by $1$.
\end{enumerate}

\medskip
It is possible to show that the winning region in $G[{\geq}c]$ is
$\fattr{X}$.
Indeed, every node in the winning region of $G[{\geq}c]$ can be
attracted to $X$ without passing through colors smaller than $c$
infinitely often.
In the other direction, the attractor strategy to $X$ induced by
$\fattr{X}$ can be converted to a winning strategy in $G[{\geq}c]$. 
The size of $G[{\geq}c]$ is bounded by the size of $G$: there is at
most one more node (the sink state), and each edge added
to $G[{\geq}c]$ has a corresponding edge that is removed from $G$.

\item As before, the computation of {\tt layeredAttr}($G,p,X$) can be
completed in $O(\size V \cdot \size E)$.
Indeed, the entire run of the 
for loop can be implemented so
that each edge is crossed exactly once in all the monotone control
predecessor computations. 
Then, the loop on $X$ and the loop on $b$ can run at most $\size V
\cdot \size c$ times. 
And the number of times {\tt psolQ} is called is bounded by $\size
V$. 
\end{enumerate} 
 $\qed$\bigskip

If $\psolq$ were to restrict attention to the last two
iterations of the for loop, i.e., those that compute the greatest
fixed-point with the 
maximal even color and the maximal odd color, the run time of
$\psolq$ would be bounded by $O({\size V}^2 \cdot \size E)$. 
For such
a version of \psolq we also ran experiments on our benchmarks and do
not report these results, except to say that this version performs
considerably worse than \psolq in practice. We believe 
that this is 
so since \psolq more quickly discovers
small winning
regions that ``destabilize'' the rest of the games.

\subsection{Robustness of \psolb}
Our pseudo-code for \psolb iterates through colors in
descending order. A natural question is whether the computed output
game depends on the order in which these colors are iterated.
Notice, that for \psolq there is no such dependency.
Below, we formally state
that the outcome of \psolb~--~the residual parity game and the two
sets
$\winps {\psolb}Gp$~--~is indeed independent of the
iteration order. This suggests that the partial solver \psolb computes
a form of
polynomial-time projection of parity games onto sub-games. 

Let us formalize this.
Let $\pi$ be some sequence of colors in $G$, that may omit or repeat
some colors from $G$. 
Let $\psolb(\pi)$ be a version of \psolb that checks for (and
removes) fatal attractors according to the order in $\pi$ (including
color repetitions in $\pi$).
We say that $\psolb(\pi)$ is \emph{stable} if for every
color $c_1$, the input/output behavior of $\psolb(\pi)$ and
$\psolb(\pi\cdot c_1)$ are the same.
That is, the sequence $\pi$ leads \psolb to stabilization in the sense
that every extension of the version $\psolb(\pi)$ with one color does
not change the input/output behavior. 
\begin{theorem}
Let $\pi_1$ and $\pi_2$ be sequences of colors with
$\psolb(\pi_1)$ and $\psolb(\pi_2)$ stable. Then $G_1$ equals $G_2$ if $G_i$ is the output of $\psolb(\pi_i)$ on $G$, for $1\leq i\leq 2$.
\label{theorem:robustness}
\end{theorem}

In order to prove Theorem~\ref{theorem:robustness} we first prove a
few auxiliary lemmas. Below, we write $G[U]$ for the subgame identified by node set $U$.

\begin{lemma}
  For every game $G$, for every set of nodes $K$ and
  for every trap $U$ for player $p$, the following holds:
  $\attr{G}{K} \cap U \subseteq \attr{G[U]}{K\cap U}$
  \label{lemma:attractors in traps are larger}
\end{lemma}

\beginproof
  The proof proceeds by induction on the distance from $K$ in
  $\attr GK$. For every node $v$ of $G$ let $d(v)$ denote the
  distance of $v$ from $K$ in the attraction to $K$ in $G$.

\begin{itemize}
\item  Suppose that $K\cap U=\emptyset$.
  Then, $\attr{G[U]}{K\cap U} =\emptyset$ and we have to show
  that $\attr GK \cap U = \emptyset$.

  Assume otherwise, then $v\in \attr GK \cap U \neq \emptyset$.
  Let $v$ be the node of minimal distance to $K$ in $\attr GK \cap U$.
  If $v\in V_p$, then there is some successor $w$ of $v$ such that
  $d(v)=d(w)+1$.
  However, $w$ cannot be in $\attr GK \cap U$ by minimality of $v$.
  Thus, there is an edge from $v$ that leads to a node not in $U$
  contradicting that $U$ is a trap for player $p$.
  Similarly, if $v\in V_{1-p}$, then for all successors $w$ of $v$ we have 
  $d(v)>d(w)$ and it follows that all successors $w$ of $v$ are not in
  $\attr GK \cap U$.
  So all successors of $v$ are not in $U$ and $U$ cannot be a trap for
  player $p$.

  It follows that $\attr GK \cap U=\emptyset$ as required.

\item   Suppose that $K\cap U \neq \emptyset$.
  We prove that for every node $v\in \attr GK \cap U$ we have 
  $d_G(v,K) \geq d_{G[U]}(v,K\cap U)$, where $d_G(v,K)$ and
  $d_{G[U]}(v,K\cap U)$ are the distances of $v$ from $K$ (respectively $K\cap
  U$) in the computation of the corresponding attractor.

  Again, the proof proceeds by induction on $d_G(v,K)$.
  Consider a node $v$ in $\attr GK \cap U$ such that $d_G(v,K)=0$.
  Then $v$ is in $K$ and from $v\in U$ we conclude that $v$ is in $K\cap U$ and
  $d_{G[U]}(v,K\cap U)=0$.

  Consider a node $v$ in $\attr GK \cap U$ such that $d_G(v,K)>0$.
  If $v$ is in $V_p$, then there is a node $w$ such that $d_G(v,K)=d_G(w,K)+1$.
  Since $U$ is a trap, it must be the case that $w$ is in $U$ as well and hence
  $w$ is in $\attr GK \cap U$. By induction $d_G(w,K) \geq d_{G[U]}(w,K\cap U)$.

  If $v$ is in $V_{1-p}$, then for all successors $w$ of $v$ we have
  $d_G(v,K)\geq d_G(w,K)+1$. 
  Furthermore by $U$ being a trap, there is some successor $w$ of $v$ such
  that $w$ is in $U$.
  It follows that $w$ is in $\attr GK \cap U$.
  
  As $U$ is a subset of the nodes of $G$ we have 
  $succ(v,G)\supseteq succ(v,G[U])$, where $succ(v,G)$ is the set of
  successors of $v$ in $G$ and $succ(v,G[U])$ is the set of successors
  of $v$ in $G[U]$.
  But then, for every $w$ in $succ(v,G[U])$ we have
  $d_{G[U]}(w,K\cap U) \leq d_{G}(w,K)$.
  Hence, $d_{G[U]}(v,K\cap U) \leq d_{G}(v,K)$.
\end{itemize}
 $\qed$\bigskip

We now specialize the above to the case of monotone attractors.
We narrow the scope in this context to match its usage in $\psolb$. 
A more general claim talking about general sets in the spirit of
Lemma~\ref{lemma:attractors in traps are larger} requires quite
cumbersome notations and we skip it here (as it is not needed below).

\begin{lemma}
  Consider a game $G$ and a set of nodes $K$ of color $c$ such that
  $p=c\%2$. 
  For every trap $U$ for player $p$, the following holds:
  $\nattr{\emptyset}{K}c \cap U$ computed in $G$ is a subset of
  $\nattr{\emptyset}{K\cap U}c$ computed in $G[U]$. 
  \label{lemma:monotone attractors in traps are larger}
\end{lemma}

\beginproof
The proof is very similar to the proof of Lemma~\ref{lemma:attractors
  in traps are larger} and proceeds by induction on the distance from $K$ in
  $\nattr \emptyset Kc$. For every node $v$ of $G$ let $d(v)$ denote the
  distance of $v$ from $K$ in the monotone attraction to target $K$ in
  $G$.

\begin{itemize}
\item  Suppose that $K\cap U=\emptyset$.
  Then, $\nattr{\emptyset}{K\cap U}c$ in $G[U]$ is empty and we have
  to show that $\nattr \emptyset Kc$ in $G$ has empty intersection
  with $U$.

  Assume otherwise, then there is some $v$ such that $v$ is in $\nattr
  \emptyset Kc$ in $G$ and $v\in U$.
  Let $v$ in $U$ be the node of minimal distance to $K$ in $\nattr
  \emptyset Kc$ computed in $G$.
  If $d(v)=1$ and $v\in V_p$, then $v$ has some node in $K$ as
  successor.
  But $K\cap U=\emptyset$ and $v$ has a successor outside $U$
  contradicting that $U$ is a trap.
  If $d(v)=1$ and $v$ is in $V_{1-p}$, then all successors of $v$ are in
  $K$. 
  As $K\cap U=\emptyset$ all successors of $v$ are outside $U$
  contradicting that $U$ is a trap.
  If $d(v)>1$, the case is similar.
  If $v$ is in $V_p$, then there is some successor $w$ of $v$ such that
  $d(v)=d(w)+1$.
  However, $w$ cannot be in $\nattr \emptyset Kc \cap U$ computed in
  $G$, by the minimality
  of $v$.
  Thus, there is an edge from $v$ that leads to a node not in $U$
  contradicting that $U$ is a trap for player $p$.
  Similarly, if $v$ is in $V_{1-p}$, then for all successors $w$ of $v$ we
  have $d(v)>d(w)$ and it follows that all successors $w$ of $v$ are not in
  $\nattr \emptyset Kc \cap U$ in $G$.
  So all successors of $v$ are not in $U$ and $U$ cannot be a trap for
  player $p$.

  It follows that $\nattr \emptyset Kc$ computed in $G$ does not intersect $U$
  as required.

\item  Suppose that $K\cap U \neq \emptyset$.
  We prove that for every node $v$ in $\nattr \emptyset Kc\cap U$ computed
  in $G$ we have 
  $d_G(v,K) \geq d_{G[U]}(v,K\cap U)$, where $d_G(v,K)$ and
  $d_{G[U]}(v,K\cap U)$ are the distances of $v$ from $K$ (respectively $K\cap
  U$) in the computation of the corresponding monotone
  attractors.

  Again, the proof proceeds by induction on $d_G(v,K)$.
  Consider a node $v$ in $\nattr \emptyset Kc$ computed in $G$ such
  that $v$ is in $U$ and $d_G(v,K)=1$.
  Then, if $v$ is in $V_p$, then $v$ has a successor in $K$. As $U$ is a
  trap, it must be the case that this successor is also in $U$ showing
  that $d_{G[U]}(v,K\cap U)=1$.
  If $v$ is in $V_{1-p}$, then all of $v$'s successors are in $K$.
  As $U$ is a trap, $v$ must have some successors in $G[U]$.
  It follows that $d_{G[U]}(v,K\cap U)=1$.

  Consider a node in $\nattr \emptyset Kc$ such that $v$ is in $U$ and
  $d_G(v,K)>1$. 
  If $v$ is in $V_p$ then there is a node $w$ such that $d_G(v,K)=d_G(w,K)+1$.
  By $U$ being a trap, it must be the case that $w$ is in $U$ as well and hence
  $w$ is in $\nattr \emptyset Kc\cap U$ computed in $G$. By induction
  $d_G(w,K) \geq d_{G[U]}(w,K\cap U)$.

  If $v$ is in $V_{1-p}$, then for all successors $w$ of $v$ we have
  $d_G(v,K)\geq d_G(w,K)+1$. 
  Furthermore by $U$ being a trap, there is some $w$ successor of $v$ such
  that $w$ is in $U$.
  It follows that all such $w$ are in $\nattr \emptyset Kc\cap U$
  computed in $G$.
  
  As $U$ is a subset of the nodes of $G$, we have 
  $succ(v,G)\supseteq succ(v,G[U])$, where $succ(v,G)$ is the set of
  successors of $v$ in $G$ and $succ(v,G[U])$ is the set of successors
  of $v$ in $G[U]$.
  But then, for every $w$ in $succ(v,G[U])$ we have
  $d_{G[U]}(w,K\cap U) \leq d_{G}(w,K)$.
  Hence, $d_{G[U]}(v,K\cap U) \leq d_{G}(v,K)$.
\end{itemize}
 $\qed$\bigskip

We now show that the order of removal of attractors for even and odd
colors are interchangeable.

\begin{lemma}
  Removal of fatal attractors for even colors and for odd colors are
  interchangeable. 
\label{lemma:even and odd interchangeable}
\end{lemma}

\beginproof
  Let $c_1$ be some odd color and $c_0$ be some even color.
  Let $X_1$ be the set of nodes of color $c_1$ such that $X_1\subseteq
  \nattr[1]G{X_1}{c_1}$ and $X_1$ is the maximal node set with this property.
  (That is to say, $X_1$ is the set computed by a call to \psolb with the
  color $c_1$.)
  Similarly, let $X_0$ be the set of nodes of color $c_0$ such that
  $X_0\subseteq \nattr[0]G{X_0}{c_0}$ and $X_0$ is the maximal with
  this property.
  We assume that both $\nattr[1]G{X_1}{c_1}$ and
  $\nattr[0]G{X_1}{c_1}$ are not empty. 

  By soundness, $\nattr[1]G{X_1}{c_1}$ is part of the 
  winning region for player~1. 
  Let $U$ be the residual game
  $G\setminus\attr[1]G{\nattr[1]G{X_1}{c_1}}$.
  We note that Lemma~\ref{lemma:attractors in traps are larger} does
  not help us directly.
  Indeed, node set $\attr[1]G{\nattr[1]G{X_1}{c_1}}$ is an attractor for
  player~1.
  Hence, $U$ is a trap for player~1 but not necessarily for player~0.

  By soundness, $\nattr[0] G{X_0}{c_0}$ is a subset of $U$.
  Indeed, all the nodes that are removed from $G$ are winning for
  player~1 but $\nattr[0] G{X_0}{c_0}$ is part of the winning region
  for player~0. It follows that $X_0$ is a subset of $U$.

  Furthermore, $\nattr[0]{G[U]}{X_0\cap U}{c_0}$ is a
  superset of $\nattr[0]{G}{X_0}{c_0}$, where this follows from an
  argument similar to the one made in the proof of
  Lemma~\ref{lemma:attractors in traps are larger} above.

  But from the construction of $\nattr[0]{G[U]}{X_0\cap U}{c_0}$ it
  follows that node set
  $\nattr[0]{G[U]}{X_0\cap U}{c_0}$ is also a subset of $\nattr[0]G{X_0}{c_0}$.
  Indeed, if we consider the entire doubly nested fixpoint, then the
  computation of $\nattr[0]{G[U]}{X_0\cap U}{c_0}$ starts from a
  subset of the nodes of color $c_0$ and $\nattr[0]G{X_0}{c_0}$ starts
  from the entire set of nodes of color $c_0$.
 $\qed$\bigskip

It follows that we may think about the removal of (attractors of) fatal attractors
separately for all the even colors and all the odd colors.
We now have all the tools in place to prove Theorem~\ref{theorem:robustness}:
\beginproof[of Theorem~\ref{theorem:robustness}]
  By Lemma~\ref{lemma:even and odd interchangeable}, we may assume that
  in both $\pi_1$ and $\pi_2$ all even colors occur before odd colors. 
  We show that the node set of the output of version $\psolb(\pi_1
  \cdot \pi_2)$ is a subset of the node set of the output of version
  $\psolb(\pi_2)$. As $\pi_1$ is stable, it follows that actually
  $\psolb(\pi_1) \subseteq \psolb(\pi_2)$.
  The same argument works in the other direction and it follows that
  the two residual games are actually equivalent.

  Let $\pi_1=c^1_1\cdots c^1_n$, where $c^1_1,\ldots, c^1_m$ are even
  and $c^1_{m+1},\ldots, c^1_n$ are odd.
  Let $G^1_0$, $G^1_1$, $\ldots$, $G^1_n$ be the sequence of games
  after the different applications of the colors in $\pi_1$.
  That is, $G^1_0=G$, and $G^1_{i}$ is the result of applying
  \psolb with color $c^1_i$ on $G^1_{i-1}$.
  It follows that $G^1_n=G_1$.
  Similarly, let $\pi_2=c^2_1\cdots c^2_p$, where $c^2_1,\ldots,
  c^2_q$ are even and $c^2_{q+1},\ldots, c^2_{p}$ are odd.
  Let $G^2_0=G$ and let $G^2_{i}$ be the result of applying \psolb
  with color $c^2_i$ on $G^2_{i-1}$.
  Let $G^{1,2}_0=G^1_n$ and $G^{1,2}_i$ is the result of applying
  \psolb with color $c^2_i$ on $G^{1,2}_{i-1}$.
  We show that $G^{1,2}_j$ is a subset of $G^2_j$.
  
  By Lemma~\ref{lemma:even and odd interchangeable} it is clear
  that we can consider the application of $c^2_1, \ldots, c^2_q$ right
  after the application of $c^1_1,\ldots, c^1_m$.
  Indeed, in the sequence $c^1_{m+1},\ldots, c^1_{n}$ is
  interchangeable with $c^2_1,\ldots, c^2_q$.

  Consider the application of $c^2_j$ to $G^{1,2}_{j-1}$ and to
  $G^2_{j-1}$.
  By induction $G^{1,2}_{j-1}$ is a subset of $G^2_{j-1}$.
  Furthermore, $G^{1,2}_{j-1}$ is obtained from $G$ by removing a
  sequence of attractors for player~0.
  It follows that $G^{1,2}_{j-1}$ is $G^{2}_{j-1}$ restricted to a
  trap for player~0.

  It follows from Lemmas~\ref{lemma:monotone attractors in traps are
    larger} and \ref{lemma:attractors in traps are larger} that the
  computation of the attractor removes a larger part of
  $G^{1,2}_{j-1}$ than that of $G^2_{j-1}$.
  Hence $G^{1,2}_j$ is a subset of $G^2_j$.
 $\qed$\bigskip

\subsection{Complete sub-classes for \psolb}
Next, we formally define classes of
parity games, those that \psolb\ solves completely and those that
\psolb does not modify.
We concentrate on \psolb as it seems to offer the best trade off
between efficiency and discovery (see Section~\ref{sec:results}).
\begin{definition}
We define class $\solved$ (for ``Solved'') to consist of those parity games $G$ for
which $\psolb (G)$ outputs the empty game. And we define $\kernel$ (for
``Kernel'') as
the class of those parity games $G$ for which $\psolb (G)$ outputs $G$ again.
\end{definition}

The meaning of \psolb is therefore a total, idempotent
function of type $\games\to \kernel$ that has $\solved$ as inverse
image of the empty parity game.  By virtue of
Theorem~\ref{theorem:robustness}, classes $\solved$
and $\kernel$ 
are \emph{semantic} in nature.

We now show that $\solved$ contains the
class of B\"uchi games, which we identify with parity games $G$
with color $0$ and $1$ and where nodes with color $0$ are those
that player $0$ wants to reach infinitely often.
\begin{theorem}
\label{theorem:buechi}
Let $G$ be a parity game whose colors are only $0$ and $1$. Then $G$
is in $\solved$, i.e.\ \psolb completely solves $G$.
\end{theorem}

\beginproof
We recall one way of solving a B\"uchi game, which takes the perspective
of player $0$. 
First we inductively define, for $n\geq 0$, and $X = \{v\in V\mid c(v)
= 0\}$ the sets 
\begin{eqnarray}
\label{equ:buchi}
Z^0 &=& V \qquad\qquad\ \ U^n = \attr[0]G{Z^n} \\
Z^{n+1} &=& Y^n\cap X  \qquad Y^n = \cpred [0]{U^n} \nonumber
\end{eqnarray}


Let $n_0$ be minimal such that $Z^{n_0} = Z^{n_0+1}$. The winning region for $W_0$ for player $0$ in game $G$ with colors $0$ and $1$ only is then equal to
\begin{equation}
\label{equ:W0}
W_0 = \attr[0]G{Z^{n_0}}
\end{equation}

Since the order of processing colors in \psolb does not impact its
output game (by Theorem~\ref{theorem:robustness}), we may assume that
color $d=0$ gets processed first (this is just for convenience of
presentation). 

When the first iteration of \psolb does process $d=0$, the computation
essentially captures the process defined in the
equations~(\ref{equ:buchi}): the interplay of $U^n$ and $Y^n$ achieves
the effect that player $0$ can move from $Y^n$ into $U^n$, which
models that player $0$ can reach the target set again from every node in
the target set. 
The computation of $Z^{n+1}$ corresponds to the {\tt else} branch of the
iteration within \psolb. The constraint of our monotone attractor,
that $c(v)\geq d$, is vacuously true here as $d$ equals $0$. So the
first iteration will effectively compute set $Z^{n_0}$ as
fixed-point. Then \psolb will be called recursively on $G\setminus
W_0$ by the definition of $W_0$ in~(\ref{equ:W0}). 

In that remaining game, player $1$ can secure that all plays visit
nodes of color $0$ only finitely often. This follows from the fact
that $W_0$ was removed from game $G$ and that B\"uchi games are
determined. In particular, \psolb will not detect a fatal attractor
for $d=0$ in that remaining game. But when its iteration runs with
$d=1$ we argue as follows. 

The following algorithm computes the winning region for player~1 in a
B\"uchi game. Let $X = \{ v\in V \mid c(v)=1\}$.
\begin{eqnarray}
\label{equ:cobuchi}
Z^0 &=& \emptyset \qquad\qquad\qquad\qquad\ \  Y^{n,0} = X \\
Z^n &=& \attr[1]G{Y^{n,m^n_0}} \qquad Y^{n,m} = X \cap \cpred[1]{Z^{n-1} \cup Y^{n,m-1}} \nonumber
\end{eqnarray}


\noindent where $m^n_0$ is the minimal natural number such that $Y^{n,m^n_0}$
equals $Y^{n,m^n_0+1}$.
Let $n_0$ be the minimal natural number such that $Z^{n_0}$ equals $Z^{n_0+1}$.
Let $X^{i,j}$ denote the sequence of values computed for the variable
$X$ in \psolb, where $i$ is the number of recursive invocations of
\psolb, and $j$ is the value of $X$ computed after running in the loop
$j$ times.

It is simple to see that $X^{n,m}$ is a superset of $Y^{n,m}$
restricted to the residual game in the $n$th call to \psolb.
Indeed, both start from the set $X$ and the computation of $X\cap
\cpred[1]{Z^{n-1}\cup Y^{n,m-1}}$ is contained in the computation of
$\fattr{X^{n,m-1}}$.
The intersection with $X$ in the algorithm above is included in the
definition of $\fattr{X}$.
Furthermore, every recursive call to \psolb computes the exact
attractor $\attr[1]{G}{\fattr{X}}$ just as above.
And the removal of nodes in \psolb is equivalent to the inclusion of
$Z^{n-1}$ in the computation of $\cpred[1]{Z^{n-1}\cup Y^{n,m-1}}$.
 $\qed$\bigskip

There is interest in the computational complexity of specify types of
parity games: do they have bespoke solvers that run in polynomial
time, or are they solved in polynomial time by specific general
solvers of parity games? Dittmann et al.\
\cite{Kreutzer12} prove that restricted
classes of digraph parity games can be solved in polynomial
time. Berwanger and Gr\"adel prove such polynomial run-time complexities for weak and
dull parity games \cite{DBLP:journals/mst/BerwangerG04}. Gazda and Willemse study the
behavior of Zielonka's algorithm for weak, dull, and solitaire games
and adjust Zielonka's algorithm to solve all three classes of parity
games in polynomial time \cite{DBLP:journals/corr/GazdaW13}.

It is therefore of interest to examine whether $\solved$ contains such
classes of parity games. For example, not all 1-player
parity games are in $\solved$ (see Figure~\ref{fig:pgtwo}). Since the
parity game in Figure~\ref{fig:pgtwo} is also a dull \cite{DBLP:journals/mst/BerwangerG04}
game, we infer that not all dull games are in $\solved$ either.
Class $\solved$ is also not closed under sub-games, as the next
example shows.
\begin{example}
Consider the game $G$ that is obtained from the one in Figure~\ref{fig:pg} by
adding an edge from $v_7$ to $v_3$. Then \psolb
solves this game completely as there is now also a fatal attractor for
$k=17$. But the game in Figure~\ref{fig:pg} is a sub-game of this
game whose subset of nodes $\{v_3,v_5,v_7\}$ is not solved by
\psolb~--~as discussed in Example~\ref{example:pg}.
\end{example}

We recall that a parity game $G$ is deterministic if for all its
nodes $v$, set $v.E$ has size $1$.
We record that \psolb solves completely all deterministic
games~--~the proof of this fact easily can be modified to prove the
corresponding fact for the partial solvers \psol and \psolq.

\begin{lemma}
Let $G$ be a deterministic parity game. Then \psolb solves $G$ completely.
\end{lemma}

\beginproof
Let $v$ be a node in $G$. Since $G$ is deterministic, there is exactly
one play in $G$ beginning in $v$. This play has form
$w_1w_2^\omega$ for finite words $w_1$ and $w_2$ over set $V$ and is
won by player $k\%2$ where $k$ is defined to be $\min \{ c(v')\mid
v'\in w_2\}$. Let $v'$ be in $w_2$ with $c(v') = k$.
Then the monotone attractor for $k$ in $G$
will contain at least $v'$ and so the set $X$ of nodes of color $k$ in
this attractor is non-empty. This means that $\fattr{X}$ is a fatal
attractor attractor that will be detected by \psolb~--~by virtue of
Theorem~\ref{theorem:robustness}. Since $v$ is in $\attr[k\%2]GX$, we see that
\psolb decides the winner of node $v$ in $G$.
 $\qed$\bigskip

Finally, \psolb solves all parity games that are weak in the sense of
\cite{DBLP:journals/mst/BerwangerG04}. Weak parity games $G=(V,V_0,V_1,E,c)$ satisfy that for all edges $(v,w)$
in $E$ we have $c(v)\leq c(w)$. These games correspond to model-checking problems
for the alternation-free fragment of the modal mu-calculus. The fact
that colors increase along edges means that each maximal strongly
connected component of the game graph $(V,E)$ has to have constant
color, although different components may have different colors.
We show that \psolb solves such games completely.
\begin{theorem}
Let $G$ be a parity game such that for each of its maximal strongly
connected components $C$ there is some color $c$ such that $c(v) = c$
for all $v$ in $C$. Then \psolb
completely solves $G$.
\end{theorem}

\beginproof
Let $G$ be such a game and consider the decomposition of $(V,E)$ into maximal strongly connected
components (SCCs). The set of these SCCs is a partial order with
$C\leq C'$ iff there is some $(v,w)$ in $E\cap C\times C'$. By
Theorem~\ref{theorem:robustness}, we may schedule the exploration of
colors in the execution of \psolb on $G$ in every possible order
without changing the output. Let $c$ be a color of an SCC that is maximal in the partial
order on SCCs. Then \psolb will detect a fatal attractor for $c$ that
contains $C$, and so $C$ (and possibly other nodes and edges) will be
removed from $G$. Next, \psolb will call itself recursively on this
smaller game. Since \psolb only removes normal game attractors before
making such recursive calls, we know that the remaining game also
satisfies the assumptions of this theorem.
Therefore, after we applied a new SCC decomposition on that smaller
game, we may again chose a color from some
maximal SCC that will give rise to a fatal attractor. Thus, \psolb
solves $G$ completely after at most $\size V$ many recursive calls.
 $\qed$\bigskip

%
%

\section{Experimental results}\label{sec:results}

\subsection{Experimental setup}
\label{sec:fatalexp}

We wrote Scala implementations of \psol, \psolb, and \psolq, and of
Zielonka's solver \cite{Zie98}
(\zie) that rely on the same data structures and do not
compute winning strategies~--~which has routine administrative overhead. 
The (parity) \textit{Game} object has a map of \textit{Node}s (objects) with node identifiers (integers) as the keys. Apart from colors and owner type (0 or 1), each \textit{Node} has two lists of identifiers, one for successors and one for predecessors in the game graph $(V,E)$. For attractor computation, the predecessor list is used to perform ``backward'' attraction.

This uniform use of data types allows for a first
informed evaluation. We chose \zie\ as a reference implementation
since it seems to work well in
practice on many games \cite{Friedman09}. We then compared the performance of these
implementations on all eight non-random, structured game types produced by the PGSolver tool \cite{Friedmann10}. Here is a list of brief descriptions of these game types.

\smallskip
\begin{itemize}

\item {\tt Clique}: fully connected games with alternating colors  and no self-loops.

\item {\tt Ladder}: layers of node pairs with connections between adjacent layers.

\item {\tt Recursive Ladder}: layers of $5$-node blocks with loops.

\item {\tt Strategy Impr}: worst cases for strategy improvement solvers.

\item {\tt Model Checker Ladder}: layers of $4$-node blocks.

\item {\tt Tower Of Hanoi}: captures well-known puzzle.

\item {\tt Elevator Verification}: a verification problem for an elevator model.

\item {\tt Jurdzinski}: worst cases for small progress measure solvers.
\end{itemize}

\smallskip The first seven types take as game parameter a natural
number $n$ as input, whereas
{\tt Jurdzinski} takes a pair of such numbers $n, m$ as game parameter.

For regression testing, we verified for all tested games that the winning regions of \psol, \psolb, \psolq\ and \zie\
are consistent with those computed by PGSolver.  Runs of these algorithms that took longer than
20 minutes (i.e.\ 1200K milliseconds) or for which the machine exhausted the available memory during solver computation are recorded as aborts
(``\timeout'')~--~the most frequent reason for \timeout\ was that the
used machine ran out of memory. The experiments on structured games were conducted on a machine 
with an Intel\textsuperscript{\textregistered} Core\texttrademark{} i5
(four cores) CPU at 3.20GHz and 8G of RAM, running on a Ubuntu 11.04 Linux operating system.
The random part (Section~\ref{sec:random}) and precision tuning part (Section~\ref{sec:precision}) of the experiments were conducted at a later stage, the test server used has
two Intel\textsuperscript{\textregistered} E5 CPUs, with 6-core each running at 2.5GHz and 48G of RAM.

For most game types, we used \emph{unbounded binary search}
starting with $2$ and then iteratively doubling that value,
in order to determine the \timeout\ boundary value for
parameter $n$ within an accuracy of plus/minus $10$. 
As the game type {\tt Jurdzinski}[$n,m$] has two parameters, 
we conducted three
unbounded binary searches here: one where $n$ is fixed at $10$,
another where $m$ is fixed at $10$, and a third one where $n$ equals
$m$.
We used a larger parameter configuration ($10$ $\times $ power of two) for {\tt Jurdzinski} games.

We report here only
the last two powers of two for which one of the partial solvers didn't timeout, as well as the boundary values for each solver. For game
types whose boundary value was less than $10$ ({\tt Tower Of Hanoi} and {\tt Elevator Verification}), we didn't use binary
search but incremented $n$ by $1$.
Finally, if a partial solver didn't solve its input game completely,
we ran \zie on the remaining game and added the observed running times for
\zie to that of the partial solver. (This occurred for {\tt
  Elevator Verification} for \psol and \psolb.)

\subsection{Experiments on structured games}

Our experimental results are depicted in
Figures~\ref{fig:expresults}--\ref{fig:expresultstwo}, colored
green (respectively 
red) for the partial solver with best (respectively worst)
result. Running times are reported in milliseconds. The most important
outcome is that 
partial solvers \psol and \psolb solved seven of the eight
game types \emph{completely} for all runs that did not time out, the
exception being {\tt Elevator Verification}; and that \psolq solved all
eight game types completely. This suggests that partial solvers can
actually be used as solvers on a range of structured game types.
\begin{figure}
\begin{center}
{\small
{\tt Clique}[$n$]\qquad
\hspace{26.0mm}
\begin{tabular}{|r||r|r|r|r|}
\hline
{\tt $n$}  & {\color{red}\psol} & {\color{green}\psolb} & \psolq & \zie\\ \hline\hline
2**11 & 6016.68 & {\color{red}48691.72} & {\color{green}3281.57} &  12862.92\\ \hline
2**12 & \timeout & 164126.06 & {\color{green}28122.96} & 76427.44\\ \hline
\hline
20min & ${\color{red}n=3680}$ & ${\color{green}n=5232}$ & $n=4608$  & $n=5104$\\ \hline
\end{tabular}

\vspace{3.0mm}
{\tt Ladder}[$n$]\qquad
\hspace{9.0mm}
\begin{tabular}{|r||r|r|r|r|}
\hline
{\tt $n$}  & {\color{red}\psol} & {\color{green}\psolb} & \psolq & \zie\\ \hline\hline
2**19 & \timeout & {\color{green}22440.57} & 26759.85 & 24406.79\\ \hline
2**20 & \timeout & {\color{green}47139.96} & 59238.77 &  75270.74\\ \hline
\hline
20min & ${\color{red}n=14712}$ & ${\color{green}n=1596624}$ & $n=1415776$ & $n=1242376$\\ \hline
\end{tabular}

\vspace{3.0mm}
{\tt Model Checker Ladder}[$n$]\qquad
\begin{tabular}{|r||r|r|r|r|}
\hline
{\tt $n$}  & \psol & \psolb & {\color{red}\psolq} & {\color{green}\zie}\\ \hline\hline
2**12 & {\color{red}119291.99} & 90366.80 & 117006.17 &  {\color{green}79284.72}\\ \hline
2**13 & 560002.68 & 457049.22 & {\color{red}644225.37} & {\color{green}398592.74}\\ \hline
\hline
20min & $n=11528$ & $n=12288$ & ${\color{red}n=10928}$ & ${\color{green}n=13248}$\\ \hline
\end{tabular}

\vspace{3.0mm}
\hspace{15.0mm}
{\tt Recursive Ladder}[$n$]\qquad
\begin{tabular}{|r||r|r|r|r|}
\hline
{\tt $n$}  & \psol & \psolb & {\color{green}\psolq} & {\color{red}\zie}\\ \hline\hline
2**12 & \timeout & \timeout & {\color{green}138956.08} &  \timeout\\ \hline
2**13 & \timeout & \timeout & {\color{green}606868.31} &  \timeout\\ \hline
\hline
20min & $n=1560$ & $n=2064$ & ${\color{green}n=11352}$ & ${\color{red}n=32}$\\ \hline
\end{tabular}
}
\caption{First experimental results for partial solvers run over benchmarks\label{fig:expresults}}
\end{center}
\end{figure}
\begin{figure}
\begin{center}
{\small 
\vspace{3.0mm}
{\tt Strategy Impr}[$n$]\qquad
\hspace{24.0mm}
\begin{tabular}{|r||r|r|r|r|}
\hline
{\tt $n$}  & \psol & {\color{green}\psolb} & \psolq & {\color{red}\zie}\\ \hline\hline
2**10 & 174913.85 & {\color{green}134795.46} & \timeout &  \timeout\\ \hline
2**11 & 909401.03 & {\color{green}631963.68} & \timeout &  \timeout\\ \hline
\hline
20min & $n=2368$ & ${\color{green}n=2672}$ & $n=40$  & ${\color{red}n=24}$\\ \hline
\end{tabular}

\vspace{3.0mm}
{\tt Tower Of Hanoi}[$n$]\qquad
\hspace{11.0mm}
\begin{tabular}{|r||r|r|r|r|}
\hline
{\tt $n$}  & {\color{red}\psol} & {\color{green}\psolb} & {\color{red}\psolq} & {\color{green}\zie}\\ \hline\hline
9 & 272095.32 & {\color{green}54543.31} & {\color{red}610264.18} &  56780.41\\ \hline
10 & \timeout & 397728.33 & \timeout &  {\color{green}390407.41}\\ \hline
\hline
20min & ${\color{red}n=9}$ & ${\color{green}n=10}$ & ${\color{red}n=9}$  & ${\color{green}n=10}$\\ \hline
\end{tabular}

\vspace{5.0mm}
{\tt Elevator Verification}[$n$]\qquad
\vspace{3.0mm}
\begin{tabular}{|r||r|r|r|r|}
\hline
{\tt $n$}  & {\color{red}\psol} & \psolb & \psolq & {\color{green}\zie}\\ \hline\hline
1 & {\color{red}171.63} & {\color{green}120.59} & 147.32 &  125.41\\ \hline
2 & {\color{red}646.18} & 248.56 & 385.56 &  {\color{green}237.51}\\ \hline
3 & {\color{red}2707.09} & 584.83 & 806.28 &  512.72\\ \hline
4 & {\color{red}223829.69} & 1389.10 & 2882.14 &  {\color{green}1116.85}\\ \hline
5 & \timeout & 11681.02 & 22532.75 &  {\color{green}3671.04}\\ \hline
6 & \timeout & 168217.65 & 373568.85 &  {\color{green}41344.03}\\ \hline
7 & \timeout & \timeout & \timeout & {\color{green}458938.13}\\ \hline
\hline
20min & ${\color{red}n=4}$ & $n=6$ & $n=6$ & ${\color{green}n=7}$\\ \hline
\end{tabular}
}
\caption{Second experimental results for partial solvers run over benchmarks\label{fig:expresults2}}
\end{center}
\end{figure}

\begin{figure}
\begin{center}
{\small
\vspace{3.0mm}
{\tt Jurdzinski}[$10,m$]\qquad
\hspace{1.0mm}
\begin{tabular}{|r||r|r|r|r|}
\hline
{\tt $m$}  & \psol & {\color{green}\psolb} & \psolq & {\color{red}\zie}\\ \hline\hline
10x2**7 & \timeout & {\color{green}179097.35} & \timeout &  \timeout\\ \hline
10x2**8 & \timeout & {\color{green}833509.48} & \timeout &  \timeout\\ \hline
\hline
20min & $n=560$ & ${\color{green}n=2890}$ & $n=1120$ & ${\color{red}n=480}$\\ \hline
\end{tabular}

\vspace{3.0mm}
{\tt Jurdzinski}[$n,10$]\qquad
\hspace{1.0mm}
\begin{tabular}{|r||r|r|r|r|}
\hline
{\tt $n$}  & \psol & {\color{green}\psolb} & \psolq & {\color{red}\zie}\\ \hline\hline
2**7x10 & 308033.94 & {\color{green}106453.86} & \timeout &  \timeout\\ \hline
2**8x10 & \timeout & {\color{green}406621.65} & \timeout & \timeout\\ \hline
\hline
20min & $n=2420$ & ${\color{green}n=4380}$ & $n=1240$ & ${\color{red}n=140}$\\ \hline
\end{tabular}

\vspace{3.0mm}
{\tt Jurdzinski}[$n,n$]\qquad
\hspace{6.0mm}
\begin{tabular}{|r||r|r|r|r|}
\hline
{\tt $n$}  & \psol & {\color{green}\psolb} & \psolq & {\color{red}\zie}\\ \hline\hline
2**3x10 & 215118.70 & {\color{green}23045.37} & 310665.53 & \timeout\\ \hline
2**4x10 & \timeout & {\color{green}403844.56} & \timeout &  \timeout\\ \hline
\hline
20min & $n=110$ & ${\color{green}n=200}$ & $n=100$ &  ${\color{red}n=50}$\\ \hline
\end{tabular}
}
\caption{Third experimental results run over {\tt
    Jurdzinski} benchmarks\label{fig:expresultstwo}}
\end{center}
\end{figure}

We now compare the performance of these partial solvers and of \zie.
There were ten experiments, three for {\tt Jurdzinski} and one for
each of the
remaining seven game types.


For seven out of these ten experiments, \psolb had the largest
boundary value of the parameter and so seems to perform best
overall. The solver \zie was best for {\tt Model Checker Ladder} and
{\tt Elevator Verification}, and about as good as \psolb for {\tt
  Tower Of Hanoi}. And \psolq was best for {\tt Recursive
  Ladder}. Thus \psol appears to perform worst
across these benchmarks. 

Solvers \psolb and \zie seem to do about equally well for game types {\tt Clique}, {\tt Ladder}, {\tt Model Checker Ladder}, and
{\tt Tower Of Hanoi}. But solver \psolb appears to
outperform \zie dramatically for game types
{\tt Recursive Ladder}, and {\tt Strategy Impr} and is considerably
better than \zie for game type {\tt Jurdzinski}.
 
We think these results are encouraging and 
corroborate that partial solvers based on fatal attractors
may be components of faster solvers for parity games.

\subsection{Number of detected fatal attractors}
We also recorded the number of fatal attractors that were detected in
runs of our partial solvers. One reason for doing this is to see
whether game types have a typical number of dynamically detected fatal
attractors that result in the complete solving of these games. 

We report these findings for \psol and \psolb first: for {\tt Clique},
{\tt Ladder}, and {\tt Strategy Impr} these games are solved by
detecting two fatal attractors only; {\tt Model Checker Ladder} was
solved by detecting one fatal attractor. For the other game types \psol and
\psolb behaved differently. For {\tt Recursive Ladder}[$n$], \psolb
requires $n=2^k$ fatal attractors whereas \psolq needs only
$2^{k-2}$ fatal attractors. For {\tt Jurdzinski}[$n, m$], \psolb
detects $mn+1$ many fatal attractors, and \psol removes $x$ edges
where $x$ is about $nm/2
\leq x\leq nm$, and detects slightly more than these $x$
fatal attractors. Finally, for {\tt Tower Of Hanoi}[$n$], \psol
requires the detection of $3^n$ fatal attractors whereas \psolb solves
these games with detecting two fatal attractors only.

We also counted the number of recursive calls for \psolq: it equals
the number of fatal attractors detected by \psolb for all game types
except {\tt Recursive Ladder}, where it is
$2^{k-1}$ when $n$ equals $2^k$.

\subsection{Experiments on variants of partial solvers}
We performed additional experiments on variants of these partial
solvers. Here, we report results and insights on two such
variants. The first variant is one that modifies the definition of the
monotone control predecessor to
\begin{eqnarray}
\npred{A}{X}{c} &=& 
\{ v\in V_{p}\mid ((c(v)\% 2 = p) \lor c(v)\geq c) \land v.E\cap (A\cup
X)\neq \emptyset\} \cup  \nonumber \\
& & 
\{v\in V_{1-p}\mid ((c(v)\% 2 = p)\lor c(v) \geq c) \land v.E\subseteq A\cup
  X\}\nonumber
\end{eqnarray}
\noindent The change is that the constraint $c(v)\geq c$ is weakened
to a disjunction $(c(v)\% 2 = p) \lor (c(v)\geq c)$ so that it suffices
if the color at node $v$ has parity $p$ even though it may be smaller
than $c$.  This implicitly changes the definition of the
monotone attractor and so of all partial solvers that make use of this
attractor; and it also impacts the computation of $A$ within \psolq. 
Yet, this change did not have a dramatic effect on our partial
solvers. On our benchmarks, the change 
improved things
slightly for \psol and made it slightly worse for \psolb and \psolq.

A second variant we studied was a version of \psol that removes at
most one edge in each iteration (as opposed to all edges as stated in Fig.~\ref{fig:partialsolver}).
For games of type {\tt Ladder}, e.g., this variant did much worse. But
for game types {\tt Model Checker Ladder} and {\tt Strategy Impr}, this variant did much better.
The partial solvers based on such variants and their combination are
such that \psolb (as defined in Figure~\ref{fig:buchi}) is still better across all benchmarks.

\subsection{Experiments on random games}
\label{sec:random}
With \psolb having the best overall behavior over the structured
games, we proceed to check its behavior over random games.
It is our belief that comparing the behavior of parity game solvers on 
random games does not give an impression of how these solvers perform 
on parity games in practice. 
However, evaluating how often \psolb completely solves random games complements
the insight gained above that it completely solves many structured
types of games. The experiment we conducted for this evaluation generated $100{,}000$ games 
with the {\tt randomgame} command of PGSolver for each of $16$
different configurations, rendering a total of $1.6$ million games
for that experiment.
All of these games had $500$ nodes and no self-loops. A configuration
had two parameters: a pair $(l,u)$ of minimal out-degree $l$ and
maximal out-degree $u$ for all nodes in the game (ranging over $(1,
5)$, $(5, 10)$, $(50, 250)$, and $(1, 100)$ and where the out-degree
for each node is chosen at random within the integer interval $[l,u]$), and a
bound $c$ on the number
of colors in the game (ranging over $500$, $250$, $50$, and $5$ and
where colors at nodes are chosen at random).

This gave us $4\cdot 4 = 16$ configurations for random games on which
we ran \psolb. 
The results are shown in Figure~\ref{fig:psolBresults}.
From the results in that figure we see that the behavior of \psolb was similar across the four
different color bounds for each of the four out-degree pairs
$(l,u)$. For sake of brevity, we therefore only discuss here its behavior
in terms of those out-degree pairs. Our results 
show that \psolb did not solve completely only $4{,}534$ of all $1.6$
million random games
($99.9972\%$ solved completely).
Breaking this down further, we see that when the edge density is low,
with out-degree pair  $(1,5)$, \psolb\ did not solve completely only $4{,}529$ of the
corresponding $400{,}000$ random games ($99.9887\%$ solved completely). 
The percentage of completely solved games increased to $99.9999875\%$ for the $400{,}000$ games with
out-degree pair $(5,10)$ as only $5$ of these games were then not solved
completely. For the remaining $800{,}000$ games, those with
out-degree pairs $(50, 250)$ or $(1, 100)$, all were completely solved by \psolb,
i.e.\ it solved $100$\% of those games.
The average \psolb\ run-time over these $1.6$ million games was $22$ms.
\begin{figure}
\begin{center}
{\small
{
\begin{tabular}{|l|l|r|r|}
\hline
$c$ & $(l,u)$ & \# nonempty & runtime  \\ \hline\hline
500 & (1,5) & 1086 & 22.56\\ \hline
250 & (1,5) & 1138 & 21.04 \\ \hline
50 & (1,5) & 1030 & 20.79  \\ \hline
5 & (1,5) & 1275 & 21.40 \\ \hline
\hline
500 & (5,10) & 2 & 13.08 \\ \hline
250 & (5,10) & 2 & 13.21 \\ \hline
50 & (5,10) & 1 & 12.93 \\ \hline
5 & (5,10) & 0 & 14.72 \\ \hline
\end{tabular}

\bigskip
\bigskip
\begin{tabular}{|l|l|r|r|}
\hline
$c$ & $(l,u)$ & \# nonempty & runtime  \\ \hline\hline
500 & (50,250) & 0 & 38.63 \\ \hline
250 & (50,250) & 0 & 39.07 \\ \hline
50 & (50,250) & 0 & 41.35 \\ \hline
5 & (50,250) & 0 & 37.15 \\ \hline
\hline
500& (1,100) & 0 & 17.04 \\ \hline
250 & (1,100) & 0 & 17.01 \\ \hline
50 & (1,100) & 0 & 17.69 \\ \hline
5 & (1,100) & 0 & 23.69 \\ \hline
\end{tabular}
}
}
\end{center}
\caption{Experimental results for \psolb.
  Each row shows for $100{,}000$ random games $G$ with color bound
  $c$ and out-degree pair $(l,u)$: how often \psolb
  did not solve games completely (\# nonempty), and average
  running times in milliseconds of \psolb. All but $4{,}534$ of these
  $1.6$ million games were solved completely} \label{fig:psolBresults}
\end{figure}


%

\section{Tuning the precision of partial solvers}
\label{sec:transformation}

So far, we constructed partial solvers that result from variants of
monotone attractor definitions and that simply remove such attractors
whenever they are fatal. We now suggest another principle for building
partial solvers, one that takes a partial solver as input and outputs
another partial solver that may increase the precision of its input
solver. 
As before, we concentrate on \psolb as it seems to offer the best
balance of performance and accuracy.

\subsection{Partial solver transformation $\lift$}
We fix notation for removing choices from a parity game:
\begin{definition}
Let $G = (V,V_0,V_1,E,c)$ be a parity game and $e=(v,w)$ an edge in
$E$. 
\begin{enumerate}
\item Parity game $G_e$ equals $(V,V_0,V_1,E',c)$ with $E' = \{(v',w')\in E\mid v'\not= v \hbox{ or } w' = w\}$.

\item Parity game $G\setminus e$ equals $(V,V_0,V_1,E\setminus \{e\},c)$.
\end{enumerate}
\end{definition}

Game $G_e$ is obtained from game $G$ by selecting an edge in $G$ and then
removing all edges from the source of $e$ that do not point to its
target node. This makes the game deterministic at the source node of
$e$. And game $G\setminus e$ simply removes edge $e$ from $G$.
We next introduce formal properties of partial solvers that are useful for
reasoning about the transformation of partial solvers that we will
define below.
\begin{definition}
Let $\parts$ be a partial solver.
\begin{enumerate}
\item {\bf Soundness:} $\parts$ is sound if for all games $G$ all nodes in $\winps {\parts}G0$ are won by player $0$ in $G$, and all nodes
  in $\winps {\parts}G1$ are won by player $1$ in $G$.


\item {\bf Idempotency:} $\parts$ is idempotent if for all games $G$
  as input game, the output games for $\parts$ 
  and the sequential composition of $\parts$ with itself are equal:
  $\rho(G) = \rho(\rho(G))$. 

\item {\bf Locality:} $\parts$ is local if for all games $G$, 
all players $p$ in $\{0,1\}$, and all edges $e=(v,w)$ in $G$ with
$v$ in $V_p$ we have that 
$\winps {\parts}{G}{1-p} =\emptyset$ and 
$\winps {\parts}{G_e}{1-p} \not= \emptyset$ 
imply $v\in \winps {\parts}{G_e}{1-p}$.
\end{enumerate}
\end{definition}

The property {\bf Locality} considers scenarios in which partial
solver $\rho$ cannot decide winning nodes for player $1-p$ in game
$G$, but where $\rho$ can decide winning nodes for player $1-p$ in $G$
after we restrict node $v$ in $V_p$ to have $w$ as only successor in
the game graph. In such scenarios, locality of $\rho$ means that
$\rho$ then also decides node $v$ to be won by player $1-p$ in the
restricted game $G_{(v,w)}$. This behavior is expected, for example,
when a partial solver decides winning nodes through a variant of
attractor computations as studied in this paper. 
We formally prove that \psolb satisfies these properties.
\begin{lemma}
Partial solver \psolb satisfies {\bf Soundness}, {\bf Idempotency}, and {\bf Locality}.
\end{lemma}

\beginproof
{\bf Soundness} of \psolb has been shown in Theorem~\ref{theorem:psol
  soundness}. {\bf Idempotency} follows from
Theorem~\ref{theorem:robustness}. We now show {\bf Locality}. Let $G$,
$p$, and $e=(v,w)$ in $E_G$ be given with $v\in V_p$ such that
$\winps {\psolb}{G}{1-p} =\emptyset$
and 
$\winps {\psolb}{G_e}{1-p} \not= \emptyset$. 
Proof by contradiction: Assume
that $v$ is not in $\winps {\psolb}{G_e}{1-p}$. Then $w$ also cannot
be in $\winps {\psolb}{G_e}{1-p}$ since $v.E_{G_e} = \{w\}$ and since
$\winps {\psolb}{G_e}{1-p}$ is a $1-p$ attractor in game $G_e$ by
definition of \psolb. But then neither $v$, nor $w$, nor the edge
$(v,w)$ can be part of a fatal attractor discovered in $\winps
{\psolb}{G_e}{1-p}$. From $\winps {\psolb}{G_e}{1-p} \not= \emptyset$
we know that the run $\psolb(G_e)$ discovers at least one such fatal
attractor. But since neither $v$ nor $w$ are contained in it, this
would also be a fatal attractor in $G$, contradicting that $\winps
{\psolb}{G}{1-p} =\emptyset$. 
 $\qed$\bigskip

We now describe a transformation of partial solvers that is sound for partial solvers that satisfy the above properties. Pseudo-code for our transformation of partial solvers is depicted in
Figure~\ref{fig:transformation}. Function $\lift$ takes a partial
solver $\parts$ as input and outputs another partial solver $\lift(\parts)$.
The pseudo-code describes the behavior of $\lift(\parts)$ on a parity game
$G$.

The partial solver $\lift(\parts)$ first applies partial solver $\parts$ to game
$G$ and resets $G$ to the sub-game of $G$ of nodes that $\parts$ did not
decide to be won by some player. Next, an iteration starts over all
nodes of the remaining game that have out-degree $> 1$. For such a node $v$, we record the owner of $v$ in $p$.
We then cache in $W$ the value of node set $v.E$ and start an iteration over that node
set $W$. In each such iteration, we use
$\parts$ to compute a winning region of player $1-p$ in game
$G_{(v,w)}$. If that region is non-empty, we call $\lift(\parts)$
recursively on the game $G\setminus (v,w)$. The intuition for this is
that, by fixing the edge $(v,w)$ as the strategy of player $p$ from
node $v$ in $V_p$ makes player $p$ lose plays from that
node. Therefore, it is safe to remove edge $(v,w)$ from the game without limiting the ability of player $p$ to win node $v$.
\begin{figure}
\begin{center}
{\small
\begin{alltt}
\(\lift\)(\(\parts\))(\(G\)) \{
l1:   \(G\) = \(\parts\)(\(G\));   
l2:   for (\(v\in\!\! V\) with outdegree \(> 1\)) \{
l3:       \(p\) = owner(\(v\));
l4:       \(W\) = \(v.E\);
l5:       for (\(w\in\!\! W\)) \{
l6:           if (\(\mathsf {Win}\sb{\parts}(G\sb{(v,w)},\!\! 1-p)\) != \(\emptyset\)) \{ return \(\lift(\parts)(G\setminus\!\! (v,w))\); \}
l7:       \}
l8:   \}
l9:   return \(G\); 
\}
\end{alltt}
}
\end{center}
\caption{Pseudo-code of a transformation that takes a partial solver
  $\parts$ of parity games as input and returns another partial solver.
\label{fig:transformation}
}
\end{figure}

We emphasize that $\lift(\parts)$ does not \emph{directly} detect winning
nodes, it merely removes edges. Rather, the detection of winning nodes is done
by $\parts$ itself at program point ${\tt l1}$. We illustrate this
with an example.
\begin{example}
Reconsider the parity game $G$ in Figure~\ref{fig:pgtwo} and the following
execution of $\lift(\psolb)(G)$: the initial assignment to $G$ won't
change $G$ as \psolb cannot detect winning nodes in $G$. Suppose
that the execution first picks $v$ to be $v_4$ and chooses as first $w$
node $v_3$. Then both $v_3$
and $v_4$ are contained in $\winps {\psolb}{G_{(v_4,v_3)}}0$, which is
  therefore non-empty. The execution therefore removes edge
  $(v_4,v_3)$ from $G$, calls \psolb on the resulting game, and
  assigns its output to $G$. But that output is the empty game since
  the removed edge make the nodes of color $0$ a fatal attractor for
  that color. We conclude that $\lift(\psolb)$ solves this game
  completely.
\end{example}

We now analyze the computational overhead of $\lift$ when called with a
partial solver $\parts$,
and prove that $\lift(\parts)$ is sound for partial solvers $\parts$
that satisfy the above formal properties.
\begin{theorem}
\label{theorem:lift}
Let $\parts$ be a sound partial solver.
Then we have the following:
\begin{enumerate}
\item If partial solver $\parts$ satisfies {\bf Soundness}, 
{\bf Itempotency}, and {\bf Locality} then $\lift(\parts)$ satisfies
  {\bf Soundness}.

\item The computational time complexity of $\lift(\parts)$ is in
  $O((\size E - \size V)^2\cdot \size
  \parts))$ where $\size {\parts}$ is the computational time complexity of partial
  solver $\parts$. In particular, $\lift(\parts)$ runs in polynomial time if $\parts$
  does.
\end{enumerate}
\end{theorem}

\beginproof
Let $\rho$ be a partial solver that satisfies these properties. Let $G$ be a parity game.
\begin{enumerate}
\item Consider the run of $\lift(\parts)(G)$. Each execution of its body removes a (possibly empty) node set $X_i$ from the game at program point ${\tt l1}$. If a recursive call happens (in the if-branch at program point ${\tt l6}$), this node removal event $X_i$ is then followed by the removal of an edge $e_i$ from the game. We can therefore capture essential state change information for such a run by a finite sequence
\begin{equation}
\label{equ:run}
X_0, e_0, X_1, e_1, \dots, X_{m-1}, e_{m-1}, X_m
\end{equation}

\noindent where the removal of node set $X_m$ results in a game that
is the output of $\lift(\parts)(G)$ (no more recursive calls occur
thereafter). Since $\parts$ satisfies {\bf Soundness}, we can conclude
that the decisions of winning nodes made implicitly by
$\lift(\parts)(G)$ in node sets $X_i$ are sound provided that the edge
removals in the above sequence change neither winning regions nor the
sets of winning strategies in these games. We formalize the latter
notion now: 

Let $G$ and $G'$ be parity games that have the same set of nodes and
the same coloring function. We write $G\equiv G'$ iff the winning
regions of these parity games are equal and the sets of winning
strategies of players, when restricted to their winning regions, are
equal as well. So let $G$ be the state of the game in the run
in~(\ref{equ:run}) right before edge $e_i$ gets removed. And let $G'$
be $G\setminus e_i$. It suffices to show that $G\equiv G'$ since then
all edge removals performed in~(\ref{equ:run}) preserve winning
regions and sets of winning strategies until the next set of nodes
$X_j$ gets removed from the game. Soundness of $\parts$ and the
transitivity of $\equiv$ then guarantee that decisions made implicitly
by the sound partial solver $\parts$ in node set $X_j$ are sound as
well. 

We now prove that $G\equiv G\setminus e_i$ where $e_i = (v,w)$ and $v$ in $V_p$. We do a case analysis over which player wins node $v$ in $G$:
\begin{itemize}
\item Let $v$ be won by player $1-p$ in (the current state of)
  $G$. Since $v$ is owned by player $p$ and since $G'$ equals
  $G\setminus (v,w)$ we infer $G\equiv G'$: node $v$ is not in the
  winning region of player $p$ and so winning strategies of player $p$
  won't differ when restricted to the winning region of player
  $p$. Hence the sets of winning strategies for both players,
  restricted to their winning regions, are equal in $G$ and in
  $G'$. And their winning regions are also equal since player $1-p$
  wins node $v$ owned by player $p$ and so the edge chosen at $v$
  affects the winning status of no nodes. So $G\equiv G'$
  follows. 

\item Let $v$ be won by player $p$ in (the current state of) $G$. Let
  $\sigma$ be a winning strategy of player $p$ on her winning region
  in $G$. Then $\sigma(v)$ is defined on that winning region. Proof by
  contradiction: let $\sigma(v) = w$. Since $(v,w)$ gets removed from
  the current $G$ in this run, we know that 
\begin{equation}
\label{equ:local}
\winps {\psolb}{G_{(v,w)}}{1-p} \not= \emptyset\hbox{ and }\winps {\psolb}{G}{1-p} = \emptyset
\end{equation}

\noindent where the latter is true since program point ${\tt l1}$ got
executed and since $\parts$ satisfies {\bf Idempotency}. Since
$\parts$ also satisfies {\bf Locality}, we infer
from~(\ref{equ:local}) that $v$ is in $\winps
{\psolb}{G_{(v,w)}}{1-p}$. Since $\parts$ satisfies {\bf Soundness},
we conclude from this that $v$ is won by player $1-p$ in
$G_{(v,w)}$. Since $\sigma(v) = w$ and $\sigma$ is a winning strategy
for player $p$, we know that $G$ and $G_{(v,w)}$ have the same winning
regions. Therefore, we know that $v$ is also won by player $1-p$ in
game $G$. But this is a contradiction to this second case. Thus, we
know that $\sigma(v)\not=w$. In particular, removing edge $(v,w)$ from
$G$ won't change the sets of winning regions of either player and it
won't change the sets of winning strategies for either player. So we
showed that $G\equiv G'$ holds. 

\end{itemize}

To summarize, we have shown that every sequence of edge removals
$e_i\dots e_{i+k}$ with $k\geq 0$ for which all node sets $X_{i+1}$ up
to $X_{i+k}$ are empty are such that $G\equiv G^*$ where $G$ is the
game before the removal of $e_i$, and $G^*$ is the game after removal
of $e_{i+k}$. As already discussed, this suffices to show that
$\lift(\parts)(G)$ is sound. 

\item Let $G$ be an input parity game. Let $k$ be $\size E - \size
  V$. Since each node in $G$ has out-degree at least $1$, the value
  $k$ expresses an upper bound on the number of edges that can be
  removed from $G$ in $\lift(\parts)$. 

   \begin{itemize}
   \item Let us analyze the complexity
  of the for-statement that ranges over $v\in U$: there is one initial
  call to $\parts$ and at most $k$ many calls to $\parts$ within these
  nested for-statements, and these calls are the dominating factor in
  that part of the code. Thus, an upper bound for the time complexity
  within these for-statements is $(k+1)\cdot \size\parts$. 

   \item Now we turn to the question of how often $\lift(\parts)$ may
     call itself. Each such call removes at least one edge from the
     input $G$ for the next call, and so there can be at most $k$ such calls.
   \end{itemize}

Combining this, we get as upper bound $k\cdot (k+1)\cdot \size\parts$
which is in $O(k^2\cdot \size\parts)$.
\end{enumerate}
 $\qed$\bigskip

\subsection{Experimental results for $\lift(\psolb)$}
\label{sec:precision}
We now evaluate the effectiveness of $\lift$ on the partial solver
\psolb, where we are mostly interested in the increase of precision
that $\lift(\psolb)$ has over \psolb. We evaluated this over all
structural parity game types used in earlier experiments and over the
$1.6$ million randomly generated games. As before, we applied regression
testing to confirm that all computed winning regions are consistent
with the (full) winning regions computed by PGSolver.

For the data set of $1.6$ million randomly generated games, we
ran $\lift(\psolb)$ on those $4{,}534$ games that \psolb did not solve
completely. 
%
%
Figure~\ref{fig:liftresults} shows the
results we obtained. 
Let us first discuss the results for out-degree pair $(1,5)$. Partial
solver $\lift(\psolb)$ completely solves $4{,}182$ of
the $4{,}534$ games that \psolb could not solve completely
($92.236\%$); in other words, it could not solve completely only $352$ of
these $400{,}000$ random games.  The last two columns in Figure~\ref{fig:liftresults} suggest that the run-time
overhead of $\lift(\psolb)$ over \psolb is proportional to the maximal
number of recursive calls of $\lift(\psolb)$, i.e.\ to the maximal number
of edge removals E\_rem. We saw at most $31$ such recursive calls for
these $500$ node games, and the values for N\_sol are on average about half of
the size of the node set ($500)$ of the input games.

For games of out-degree pair $(5,10)$ only five games
were not solved completely by \psolb and so only these five games
were run with $\lift(\psolb)$ for that out-degree pair. It
therefore make little sense to discuss edge and node removal and
runtimes for such
a small data set. However, we can see that  $\lift(\psolb)$ completely
solved all of these five games.

An additional result \emph{not} shown in Figure~\ref{fig:liftresults} relates to
games that are not solved completely by both \psolb and
$\lift(\psolb)$~--~a total of $347$ out of $1.6$ million.
On only four such games was $\lift(\psolb)$ able to solve additional
nodes, on the remaining $343$ games $\lift(\psolb)$ had no effect over running \psolb.
\begin{figure}
\begin{center}
{\small
{
\begin{tabular}{|l|r|r|r|r|r|r|r|}
\hline
$c$ & $(l,u)$ & \# games & \# empty & E\_rem & N\_sol & runtime \\ \hline\hline
500 & (1,5) & 1086 & 1051 & 16 & 225 & 126.90 \\ \hline
250 & (1,5) & 1138 & 1102 & 22 & 224 & 150.59 \\ \hline
50 & (1,5) & 1030 & 987 & 21 & 251 & 172.91 \\ \hline
5 & (1,5) & 1275 & 1042 & 31 & 350 & 4905.58 \\ \hline
\hline
500 & (5,10) & 2 & 2 & 1 & 4 & 19.57 \\ \hline
250 & (5,10) & 2 & 2 & 1 & 3 & 19.48 \\ \hline
50 & (5,10) & 1 & 1 & 1 & 3 & 22.33 \\ \hline
\end{tabular}
}
}
\end{center}
\caption{Experimental results for $\lift(\psolb)$ for the seven
  configurations of Figure~\ref{fig:psolBresults} in which \psolb does
  not solve all games completely: how many games \psolb could not
  solve completely (\# games), how many of those games $\lift(\psolb)$ could
  solve completely  (\# empty), the maximal number
  of edges removed in a run of $\lift(\psolb)$ (E\_rem), the maximum number of
  additional nodes solved by $\lift(\psolb)$ (N\_sol), and the average running times in milliseconds of 
  $\lift(\psolb)$} \label{fig:liftresults}
\end{figure}


For the other data set with structured parity games, we ran $\lift(\psolb)$
only over games of type ${\tt Elevator\ Verification[n]}$, since this was the only
structured game type that \psolb did not solve completely in our
experiments. In doing so, we determined that $\lift(\psolb)$ solves the same node sets that \psolb
solves, and so it also cannot solve such games completely. Therefore, we conclude that transformation $\lift$ is unable
to deal with the stuttering inherent in these games when applied to
\psolb.

\section{Conclusions}\label{sec:conclusions}
We proposed a new approach to studying the problem of solving parity
games: partial solvers as polynomial algorithms that correctly decide
the winning status of some nodes and return a sub-game of nodes for
which such status cannot be decided. We demonstrated the feasibility
of this approach both in theory and in practice. Theoretically, we
developed a new form of attractor that naturally lends itself to the
design of such partial solvers; and we proved results about the
computational complexity and semantic properties of these partial
solvers. Practically, we showed through extensive experiments that
these partial solvers can compete with extant solvers on
benchmarks~--~both in terms of practical running times and in terms of
precision in that our partial solvers completely solve such benchmark
games. 

We then suggested that such partial solvers can be subjected to a
transformation that increases their complexity within PTIME but also lets them
solve more games completely. We studied such a concrete transformation
and showed its soundness for partial solvers that satisfy reasonable
conditions. We then proved that \psolb meets these conditions and
thoroughly evaluated the effect of this transformation on \psolb over
random games, demonstrating the potential of that transformation to
increase the precision of partial solvers whilst still ensuring
polynomial time running times.

In future work, we mean to study the descriptive complexity of the
class of output games of a partial solver, for example of \psolq. We
also want to research whether such output classes can be solved by
algorithms that exploit invariants satisfied by these output
classes; insights gained in such an investigation may lead to the
design of full solvers that contain partial solvers as building blocks. Furthermore, we mean to investigate whether
classes of games characterized by structural properties of their game graphs can be solved completely by partial solvers. Such insights may connect our work to that of \cite{Kreutzer12}, where it is shown that certain classes of parity games that can be solved in PTIME are closed under operations such as the join of game graphs.
Finally, we want to investigate whether and how partial
solvers can be integrated into solver design patterns such as the one
proposed in \cite{Friedman09}. 

We refer to \cite{Huth13a} for the initial conference paper
reporting on this work, which neither contains proofs nor the material
on transforming partial solvers. A technical report \cite{Huth13}
accompanies the paper \cite{Huth13a} and contains~--~amongst other
things~--~selected proofs, the pseudo-code of our version of
Zielonka's algorithm, and further details on experimental results and
their discussion. Transformations akin to $\lift$ have been suggested already
in \cite{Wang09} as a means of making preprocessors of parity
games more effective.

\newcommand{\etalchar}[1]{$^{#1}$}

\end{document}